\documentclass[aps,prl,superscriptaddress,reprint]{revtex4-1}
\usepackage{amsmath}
\usepackage{amssymb}
\usepackage{graphicx}
\usepackage[colorlinks,urlcolor=blue]{hyperref}
\usepackage{url}
\usepackage{color,xcolor}
\usepackage{ulem}
\usepackage{float}
\usepackage{epstopdf}
\begin{document}

\title{Electronic structure, magnetic properties, spin orientation,\\ and doping effect in Mn$_3$Si$_2$Te$_6$}
\author{Yang Zhang}
\author{Ling-Fang Lin}
\affiliation{Department of Physics and Astronomy, University of Tennessee, Knoxville, Tennessee 37996, USA}
\author{Adriana Moreo}
\author{Elbio Dagotto}
\affiliation{Department of Physics and Astronomy, University of Tennessee, Knoxville, Tennessee 37996, USA}
\affiliation{Materials Science and Technology Division, Oak Ridge National Laboratory, Oak Ridge, Tennessee 37831, USA}

\date{\today}

\begin{abstract}
The layered material Mn$_3$Si$_2$Te$_6$, with alternating stacking honeycomb and triangular layers, is attracting considerable attention due to its rich physical properties. Here, using density functional theory and classical Monte Carlo (MC) methods, we systematically study this system with the $3d^5$ electronic configuration. Near the Fermi level, the states are mainly contributed by Te $5p$ orbitals hybridized with Mn $3d$ orbitals, resembling a charge transfer system. Furthermore, the spin orientations of the ferrimagnetic (FiM) ground state display different conductive behaviors when along the $ab$ plane or out-of-plane directions: insulating vs. metallic states. The energy difference between the FiM [110] insulating and FiM [001] metallic phases is very small($ \sim 0.71$ meV/Mn). Changing the angle $\theta$ of spin orientation from in-plane to out-of-plane directions, the band gaps of this system are gradually reduced, leading to an insulator-metal transition, resulting in an enhanced electrical conductivity, related to the colossal angular magnetoresistance (MR) effect. Although the three main magnetic couplings were found to be antiferromagnetic, overall the ground state is FiM. In addition, we also constructed the magnetic phase diagram using the classical $XY$ spin model studied with the MC method. Three magnetic phases were obtained including antiferromagnetic order, noncollinear spin patterns, and FiM order. Moreover, we also investigated the Se- and Ge- doping into the Mn$_3$Si$_2$Te$_6$ system: the FiM state has the lowest energy among the magnetic candidates for both Se- or Ge- doped cases. The magnetic anisotropy energy (MAE) decreases in the Se-doped case because the Mn orbital moment is reduced as the doping $x$ increases. Due to the small spin-orbital coupling effect of Se, the insulator-metal transition caused by the spin orientation disappears in the Se-doped case, resulting in an insulating phase in the FiM [001] phase. This causes a reduced colossal angular MR. However, both the MAE and band gap of the Ge-doped case do not change much with increasing doping $x$. Our results for Mn$_3$Si$_2$Te$_6$ could provide guidance to experimentalists and theorists working on this system or related materials.

\end{abstract}

\maketitle
\section{I. Introduction}

Because of their rich physical properties, layered correlated systems with transition
atoms have attracted considerable attention for decades in the condensed mater and material science communities~\cite{Imada:rmp,Moreo:science,Dagotto:science,Zhou:rmp,Cao:rpp,Varignon:nc19,Zhang:prb20,Takayama:jpsj,Do:prb22}. In those systems, many exotic physical properties are induced by
the couplings between the charge, spin, orbital, and lattice degrees of freedom, leading to colossal magnetoresistance (CMR) and electronic phase separation~\cite{Dagotto:rp,Tokura:rpp,Wu:prl01,Ward:np,Lin:prl18,miao:pnas2020}, magnetoelectricity~\cite{Sergienko:prb,Sergienko:prl,Brink:jpcm,Dong:nsr},
orbital/charge ordering~\cite{Balachandran:prb13,Varignon:prb17,Pandey:prb21,Lin:prm21}, and high-temperature superconductivity~\cite{Dagotto:rmp94,Dagotto:Rmp,Dai:rmp,Li:nature}.

\begin{figure}
\centering
\includegraphics[width=0.48\textwidth]{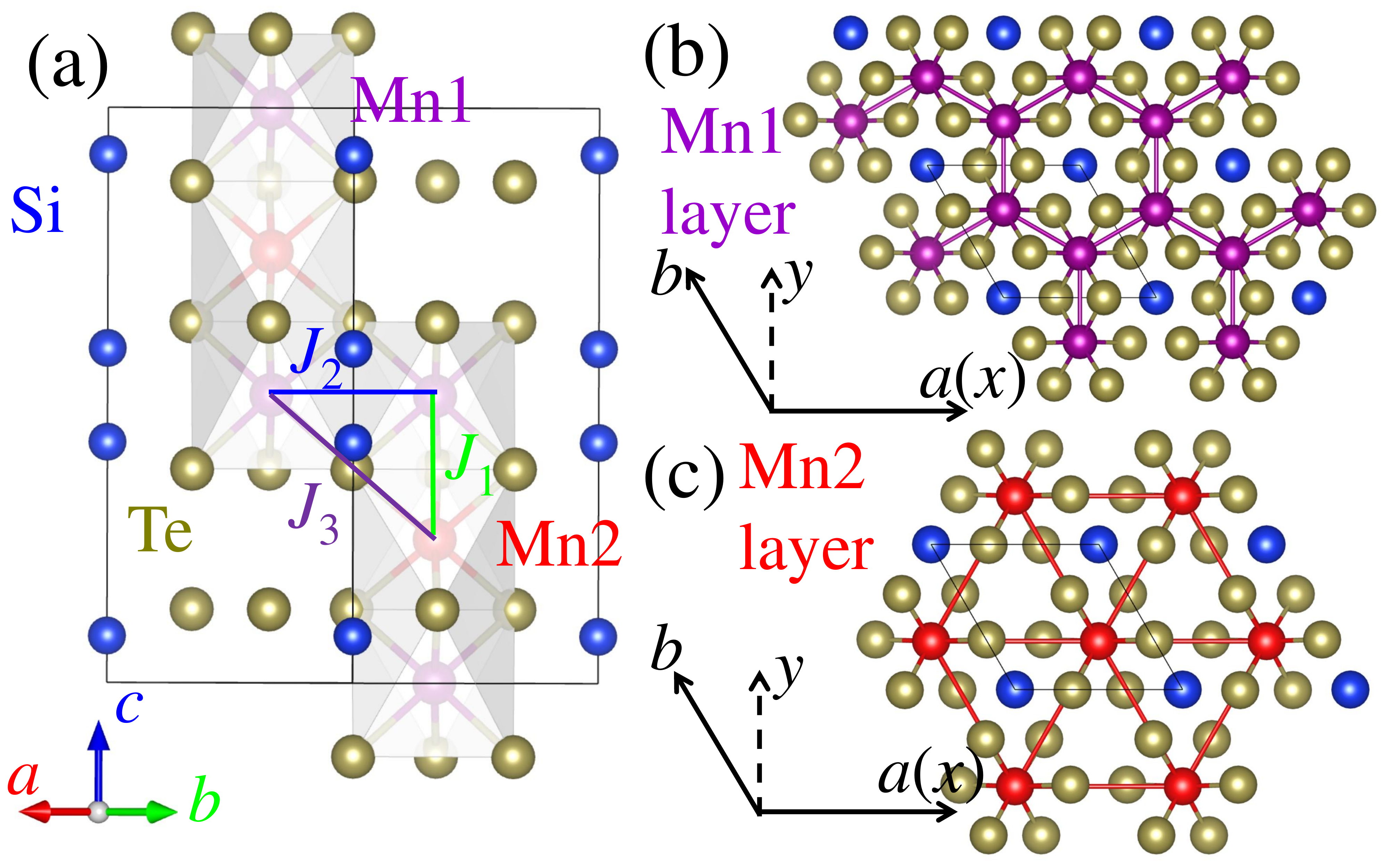}
\includegraphics[width=0.48\textwidth]{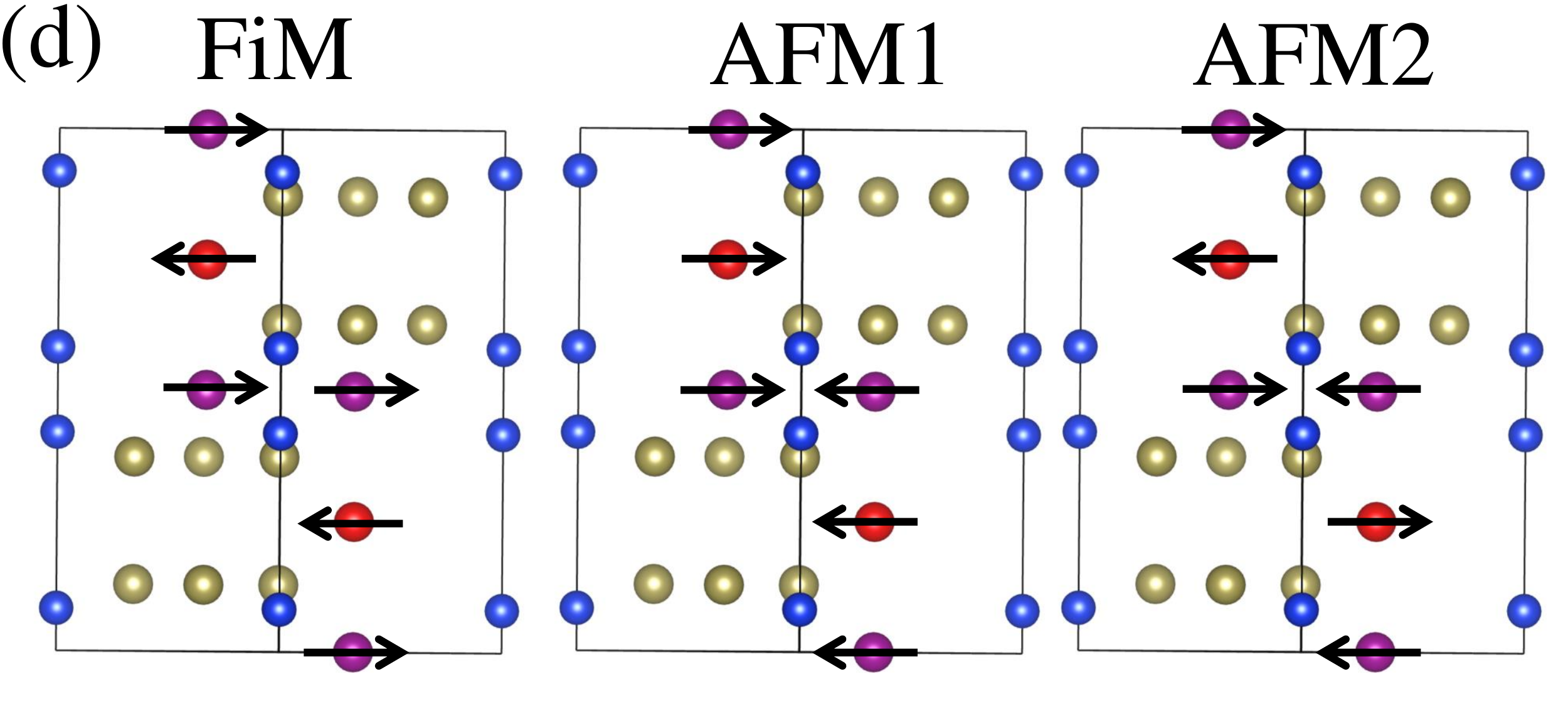}
\caption{Schematic crystal structure of Mn$_3$Si$_2$Te$_6$ with space group $P\overline{3}1c$ (No. 163). Magenta = Mn1; green = Mn2; blue = Si; dark yellow = Te. (a) Unit cell with two different Mn-atom positions, marked by different colors.
(b) Honeycomb layer of Mn1 atoms with edge sharing MnTe$_6$ octahedra along the $ab$ plane. (c) Triangular layer of Mn2 atoms with well-separated MnTe$_6$ octahedra along the $ab$ plane.
(d) Sketch of the possible magnetic orders studied here. Spin up and spin down are distinguished by the arrows.}
\label{Crystal}
\end{figure}

Recently, the layered material Mn$_3$Si$_2$Te$_6$ with mixed honeycomb and triangular layers has received considerable attention because of its interesting properties~\cite{May:prb17,Sala:prb22,Liu:prb21,Ni:prb21,Seo:nature,Wang:prb22,Zhang:nature,Ye:arxiv,Mijin:arxiv}. Mn$_3$Si$_2$Te$_6$ displays a trigonal structure with space group $P\overline{3}1c$ (No. 163), where Mn has two different atomic positions~\cite{May:prb17}, as shown in Fig.~\ref{Crystal}. The MnTe$_6$ octahedra are edge-sharing connected or are well-separated with two Mn1 and Mn2 sites in the $ab$ plane, respectively, leading to an alternately stacking of Mn-honeycomb and Mn-triangular layers along the [001] direction. The Si atoms form Si-Si dimers along the $c$-axis (see Fig.~\ref{Crystal}).

Neutron diffraction experiments found that Mn$_3$Si$_2$Te$_6$ has a ferrimagnet (FiM) order below $T_c$ $\sim 78$ K, with the spins lying along the [110] direction ($ab$ plane) due to strong anisotropy~\cite{May:prb17}. Furthermore, they also found that the largest antiferromagnetic (AFM) coupling $J_1$ is between nearest-neighbor (NN) Mn-Mn sites along the $c$-axis, while the in-plane coupling $J_2$ with the next-nearest neighbor (NNN) distance of the Mn-Mn bond is smaller than the next-next-nearest neighbor (NNNN) $J_3$. Moreover, inelastic neutron scattering was applied to the analysis of this FiM ground state in combination with the study of the spin-wave dispersion of the magnetic Hamiltonian~\cite{Sala:prb22}.

In Mn$_3$Si$_2$Te$_6$, the Mn ions are in a $2+$ valence with a $3d^5$ configuration, leading to a quenched orbital moment, resulting in a $S = 5/2$ and $L = 0$ state. Considering the half-filling $d^5$ electronic configuration of Mn$^{\rm 2+}$, the Mn should in principle favor AFM coupling, which is not the case experimentally. Hence, two simple questions naturally arise. What is the origin of the FiM order? What other interesting magnetic states can be obtained by considering the competition between the several magnetic couplings?

In addition, the CMR effect was found in Mn$_3$Si$_2$Te$_6$ by several research groups~\cite{Liu:prb21,Ni:prb21,Wang:prb22,Zhang:nature,Seo:nature}. The reported large negative CMR and thermal conductivity indicates the presence of strong spin-lattice coupling in this material~\cite{Liu:prb21}. Furthermore, the CMR occurs only when the magnetic field is applied along the magnetic hard axis and is surprisingly absent when the magnetic field is applied along the magnetic easy axis where magnetization is fully saturated ~\cite{Ni:prb21}. Meanwhile, pressure suppresses the insulating state and CMR effect in Mn$_3$Si$_2$Te$_6$, leading to a semiconductor-metal transition between $1.5$ and $2.5$ GPa, as well as a possible structural transition at $12$ GPa~\cite{Wang:prb22}. Recently, chiral orbital currents were also reported in this system~\cite{Zhang:nature}.

It was also proposed that the colossal angular magnetoresistance (MR) effect could be caused by a topological nodal-line degeneracy of spin-polarized bands in this material~\cite{Seo:nature}. Introducing carriers by doping in the system, the colossal angular MR effect slightly decreases, or is strongly suppressed at $2$~K, for the Ge- or Se- doped cases, respectively~\cite{Seo:nature}. This interesting colossal angular MR physics could be related to the rapidly reduced band gap by changing the angle $\theta$ of spin orientation between the [110] to the [001] direction. In this case, will the anisotropy change for different doping carriers? How do other physical properties, such as the magnetic ground state and anisotropic energy, change under different doping effects?
To better understand all these interesting issues, a detailed theoretical study is needed for a proper physical description of this system.

Hence, a systematic study of the physical evolution of Mn$_3$Si$_2$Te$_6$ is presented here using first-principles density functional theory (DFT) and classic Monte Carlo (MC) calculations. Based on {\it ab initio} DFT, we found that the FiM state is the most likely ground state, in agreement with neutron scattering. In addition, the states near the Fermi level are mainly contributed by the Te $5p$ states hybridized with the Mn $3d$ orbitals, leading to a charge transfer system. Furthermore, the FiM state with the spin order lying in different directions is found to display different behaviors: insulating state in the $ab$ plane and metallic state in the out-of-plane direction. Because those two phases only have a small energy difference ($\sim 0.71$ meV),  the FiM [110] insulating and FiM [001] metallic states could compete under external fields, leading to the CMR effect. By changing the angle $\theta$ between the [110] and [001] directions, the band gap is rapidly reduced, leading to an insulator-metal transition, resulting in the observed colossal angular MR effect. In addition, we also constructed the magnetic phase diagram varying magnetic couplings in a classical $xy$ spin model using the MC method. Phase competition was observed by this procedure as well.

Moreover, we also investigated the Se- or Ge-doped in the Mn$_3$Si$_2$Te$_6$ system, in the regime where the FiM state has the lowest energy among the magnetic candidates. We found that the spin still prefers to lie in the $ab$ plane for both the Se- or Ge-doped cases. As the doping $x$ level increases, the magnetic anisotropy energy (MAE) decreases in the Se-doped case due to the reduced orbital moment of Mn. Furthermore, the FiM state with spin lying in the $c$-axis displays strong insulating behavior, leading to a reduced colossal angular MR.

\section{II. Calculation Method}

\subsection{A. DFT Method}
In the present study, we employ first-principles DFT calculations performed by using the Vienna {\it ab initio} simulation package (VASP) software~\cite{Kresse:Prb,Kresse:Prb96,Blochl:Prb} with the projector augmented wave (PAW) method.
Electronic correlations were considered by using the generalized gradient approximation (GGA) with the Perdew-Burke-Ernzerhof (PBE) potential~\cite{Perdew:Prl}. The $k$-point mesh adopted was $8\times8\times4$ for the conventional cell of the $P\overline{3}1c$ structure. This $k$-point mesh was tested explicitly to confirm it produces converged energies. Furthermore, for the calculation of the density of state (DOS), the $k$-point mesh was increased to $12\times12\times6$. The plane-wave cutoff energy used was $400$~eV. Here, we considered several different collinear magnetic configurations, see Fig.~\ref{Crystal}(d). These states do not break the crystal symmetry $P\overline{3}1c$ (No. 163). In addition, on-site interactions were considered by using the local spin density approach (LSDA) plus $U_{\rm eff}$ by using the Dudarev's rotationally invariant formulation~\cite{Dudarev:prb}. Both the lattice constants and atomic positions were fully relaxed until the Hellman-Feynman force on each atom was smaller than $0.01$ eV/{\AA} for all spin configurations. All the crystal structures were visualized with the VESTA code~\cite{Momma:vesta}.

\begin{figure}
\centering
\includegraphics[width=0.48\textwidth]{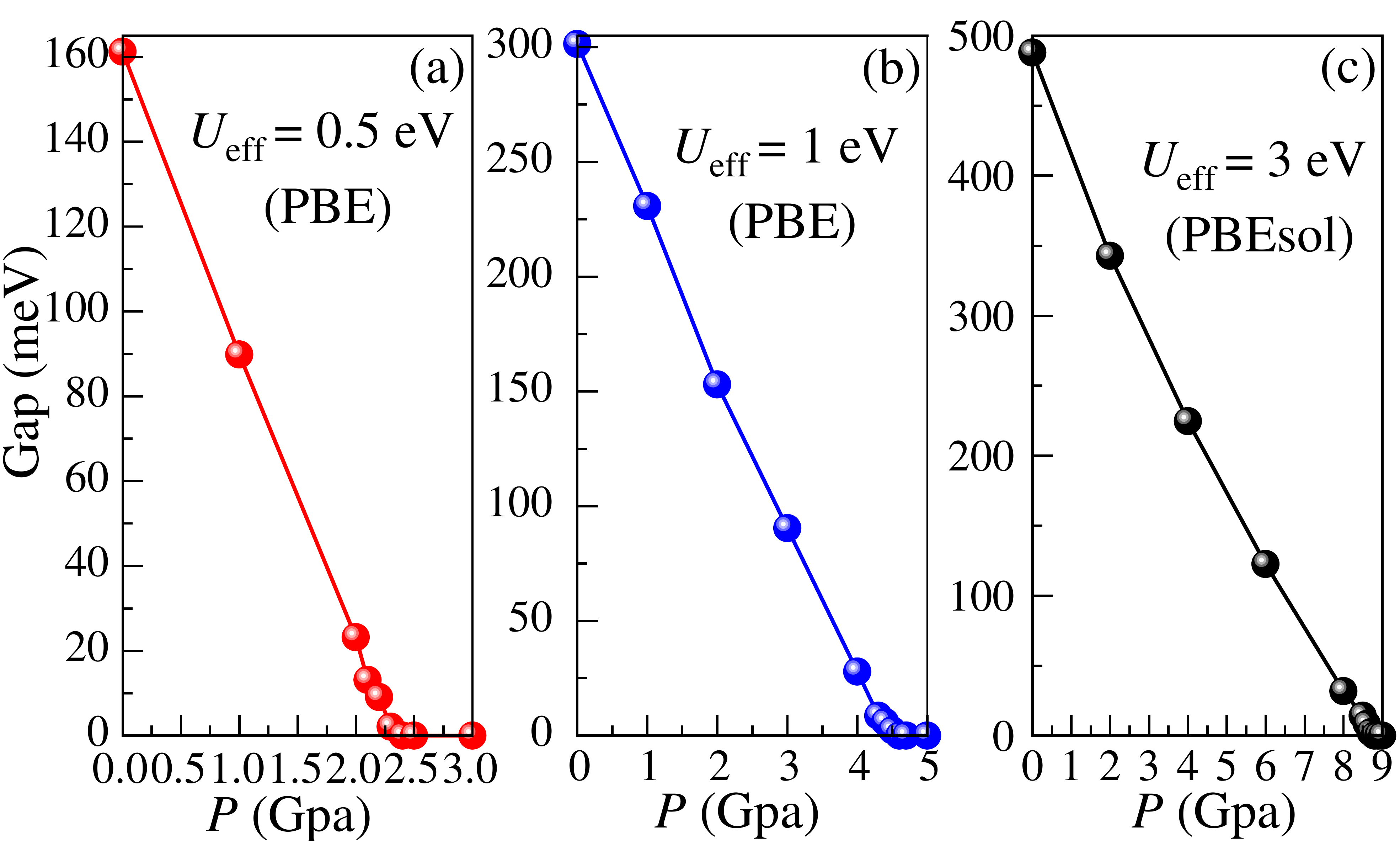}
\caption{The calculated band gap of the FiM state as a function of pressure, based on LSDA+$U_{\rm eff}$+SOC calculations. (a) $U_{\rm eff} = 0.5$ eV with PBE potential. (b) $U_{\rm eff} = 1$ eV with PBE potential. (c) $U_{\rm eff} = 3$ eV with PBEsol potential.}
\label{Pressure}
\end{figure}

Based on the $P\overline{3}1c$ (No. 163) structure of Mn$_3$Si$_2$Te$_6$, we compared the results of optimized crystal lattice constants using different values of $U_{\rm eff}$ (see Appendix). Our optimized lattice constants are $a = b = 7.058$, $c = 14.145$~\AA ~for the FiM spin state at $U_{\rm eff} = 0.5 $ eV, close to the low-temperature experimental results ($a = b = 7.017$, $c = 14.172$~\AA~\cite{Ni:prb21}). Furthermore, the pressure-induced insulator-metal transition was also experimentally observed in Mn$_3$Si$_2$Te$_6$ between $1.5$ and $2.5$ Gpa~\cite{Wang:prb22}. Based on the LSDA+$U_{\rm eff} = 0.5 $ eV calculations, we found the critical pressure for the insulator-metal transition to be about $2.4$ Gpa [see Fig.~\ref{Pressure}(a)], also close to the experimental observation ($1.5$ to $2.5$ Gpa)~\cite{Wang:prb22}. However, this critical pressure is about $4.6$ Gpa for the LSDA+$U_{\rm eff} = 1$ eV calculations, as shown in Fig.~\ref{Pressure}(b). At ambient conditions, an insulator-metal transition was also reported in Mn$_3$Si$_2$Te$_6$ by switching the magnetic field from the $ab$ plane to the $c$-axis~\cite{Seo:nature}. Our results at large $U_{\rm eff}$ could not reproduce this phase transition. For example, we obtained that the band gaps are $301.4$ and $135.9$ meV for the $ab$ plane or $c$-axis, respectively, at $U_{\rm eff} = 1 $ eV. In addition, we also tested the PBE functional revised for solids (PBEsol)~\cite{Perdew:Prl08} with $U_{\rm eff} = 3 $ eV, as used in Ref.~\cite{Wang:prb22}. The optimized lattice parameters are $a = b = 7.009$, $c = 14.058$~\AA ~and the calculated critical pressure of insulator-metal transition is $\sim 8.8$ Gpa [see Fig.~\ref{Pressure}(c)]. Furthermore, the calculated band gaps are $487.9$ and $317.7$ meV for the $ab$ plane or $c$-axis, respectively, at $U_{\rm eff} = 3 $eV with PBEsol potential. Based on these comparisons between theoretical and experimental results, then we employed the value $U_{\rm eff} = 0.5$ eV in our calculations, which is sufficient to describe this system. Overall, we conclude that the main physical results of our study are {\it not} significantly affected by the value of $U_{\rm eff}$, such as the presence of the ferrimagnetic ground state, the reduced band gap of different orientations, and the magnetic properties under doping effects. Additional discussion about the role of $U_{\rm eff}$ is presented in the Appendix and supplemental materials~\cite{Supplemental}.

\subsection{B. Monte Carlo Method}
To better understand qualitatively the magnetic properties of this system, we used a simple classic $XY$ spin model, as described below:
\begin{eqnarray}
\nonumber H&=&-J_1\sum_{<ij>}\textbf{S}_i\cdot\textbf{S}_j-J_2\sum_{[kl]}\textbf{S}_k\cdot\textbf{S}_l\\
&&-J_3\sum_{\{mn\}}\textbf{S}_m\cdot\textbf{S}_n,
\end{eqnarray}
where $J_1$, $J_2$, and $J_3$ are the exchange interactions between NN, NNN, and NNNN spin pairs, respectively, as shown in Fig.~\ref{Crystal}(a). The reason for using an $XY$ model, instead of Heisenberg model,
is because Mn$_3$Si$_2$Te$_6$ has an easy plane with the same energy in that plane.

The Markov chain MC method with the Metropolis algorithm was employed to construct
the magnetic phase diagram of this model using a $40\times40$ unit cells lattice with 6 sites in each unit cell, namely 9600 sites in total, within periodic boundary conditions.
Different lattice sizes, such as $20\times20$ and $60\times60$, gave out similar results. In the MC simulation, 1 $\times$ $10^4$ MC steps were used for thermal equilibrium at low temperature ($T = 0.05$).
For all simulated temperatures ($T$), the acceptance ratio of MC updates was kept at about $50\%$ by adjusting the
updating windows for the spin vectors to avoid being trapped in metastable states and improve the simulation efficiency~\cite{Lin:fop,Zhang:prm,Lin:prm17}.
Furthermore, the final real-space spin patterns for different parameters were obtained from the low-$T$ MC results followed by energy optimization. By this procedure, imperfections in the spin pattern were further reduced.

\section{III. Results}

\subsection{A. Magnetism and electronic structure}
Under ambient conditions, the energies of various magnetic states for the relaxed structures are summarized in Table~\ref{Table1}.
The FiM state has the lowest energy among all the candidates, in agreement with neutron experiments~\cite{May:prb17}. For the FiM state, the calculated local magnetic moments of Mn are about $4.240$ and $4.114$ $\mu_{\rm B}$/Mn for the Mn1 and Mn2 sites, respectively, corresponding to the $S = 5/2$ high-spin state of the $d^5$ Mn configuration. In addition, the calculated Mn-orbital moments are $0.031$ $\mu_{\rm B}$/Mn1 and $0.040$ $\mu_{\rm B}$/Mn2, where such a small orbital moment could induce the anisotropy. All collinear AFM states display insulating behavior with a small band gap, as shown in Table~\ref{Table1}.

\begin{table}
\centering\caption{The optimized lattice constants (\AA), calculated energy differences (meV/Mn), local magnetic moments (in $\mu_{\rm B}$/Mn) within the default PAW sphere, and band gaps (meV) for the various magnetic configurations. The FiM configuration was taken as the reference of energy. All the magnetic states discussed here were fully optimized. E[110] indicates the energy with spin lying along [110] ($ab$ plane), while the gap corresponds to the [110] direction, obtained from LSDA+$U_{\rm eff}$+SOC calculations.}
\begin{tabular*}{0.49\textwidth}{@{\extracolsep{\fill}}llllc}
\hline
\hline
                       & FiM & FM & AFM1  & AFM2 \\
\hline
$a/b$  & 7.058     & 7.068  & 7.063  & 7.066   \\
$c$    & 14.145     & 14.286 & 14.325  & 14.221   \\
E(LSDA+$U_{\rm eff}$)  & 0      & 88.54  & 14.76  & 27.91   \\
$M$(Mn1)     & 4.240  & 4.318  & 4.219 & 4.239   \\
$M$(Mn2)     & 4.114  & 4.320  & 4.182 & 4.240   \\
E[110]  & 0   & 86.67 & 14.48   &  27.05 \\
Gap & 161.2   & 0  & 561.3   & 262.6   \\
\hline
\hline
\end{tabular*}
\label{Table1}
\end{table}

As already explained, our optimized lattice constants are $a = b = 7.058$, $c = 14.145$~\AA ~for the FiM spin state, close to the experimental values at low temperature ($a = b = 7.017$, $c = 14.172$~\AA)~\cite{Ni:prb21}. In addition, the optimized in-plane lattice constants are very close for the different magnetic orders and nonmagnetic states ($a = b = 7.004$~\AA). However, the $c$-value is significantly reduced in the nonmagnetic state: $c = 11.522$~\AA, which is $2.623$~\AA ~shorter than the value of the FiM configuration. At high temperature, above the transition temperature $T_C$, no huge changes in the $c$-lattice constant were observed experimentally~\cite{May:prb17,Liu:prb21}. For this reason, short-range spin correlations should still be present above $T_c$. In fact, previous diffuse neutron scattering experiments for Mn$_3$Si$_2$Te$_6$ also revealed the existence of short-range spin correlations well above $T_C$ at 150~\cite{Ye:arxiv} and 330 K~\cite{May:prb17}, indicating possible short-range order or the persistence of correlated excitations in the paramagnetic region.

\begin{figure}
\centering
\includegraphics[width=0.48\textwidth]{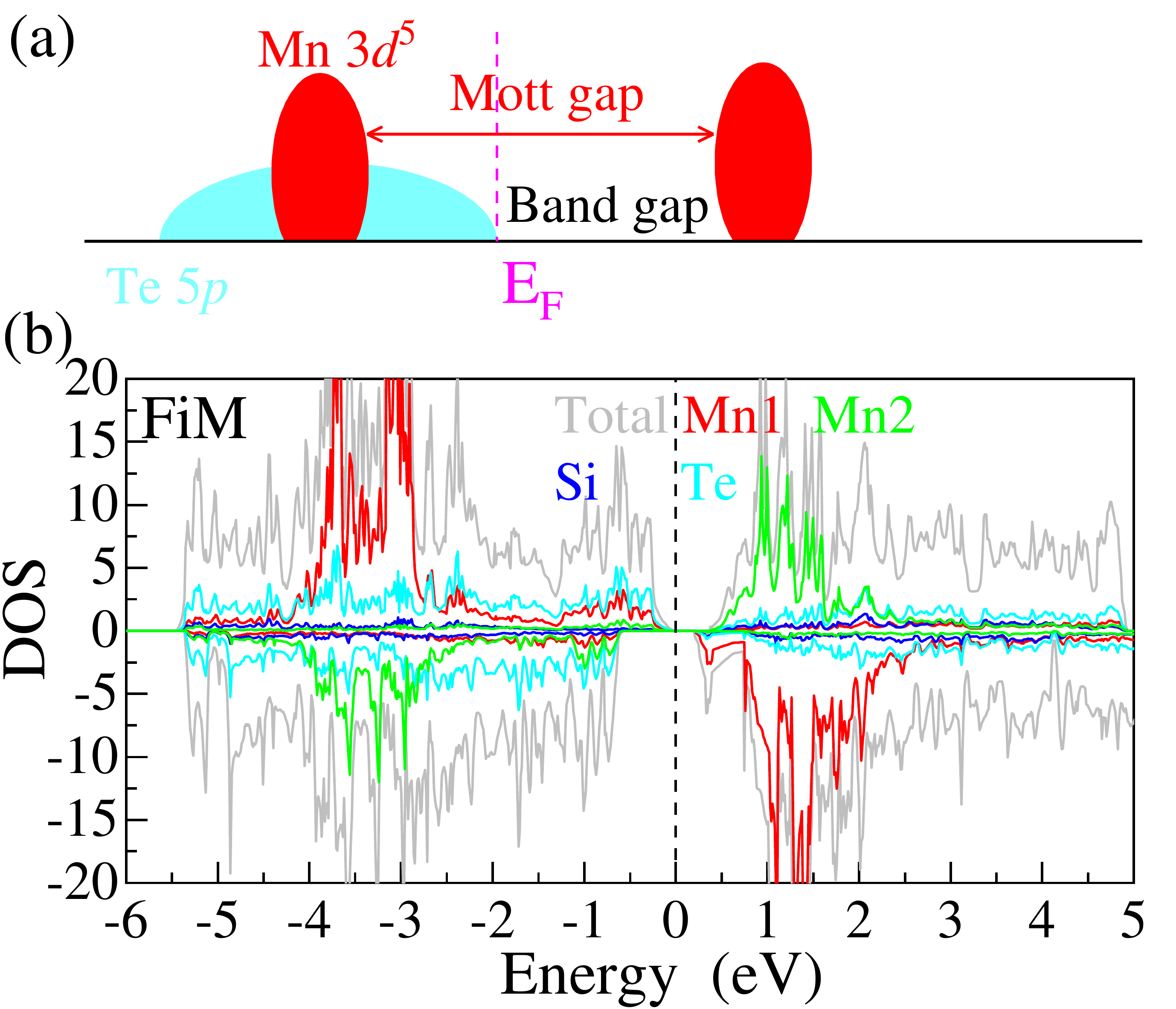}
\includegraphics[width=0.48\textwidth]{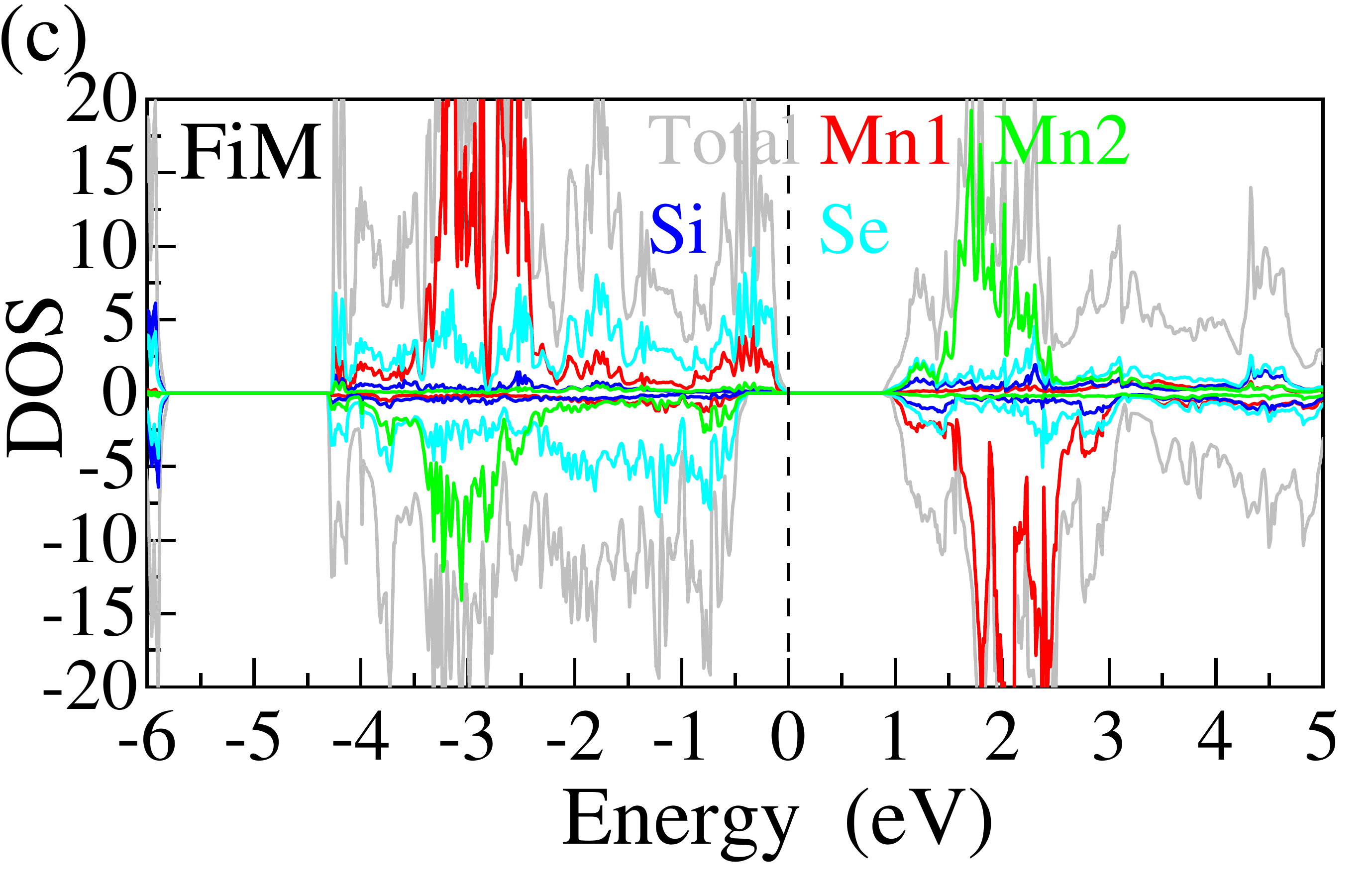}
\caption{ (a) Qualitative sketch of the local DOS for the case with Hubbard $U$ in this system that involves Mn $3d$ and Te $5p$ orbitals. (b) DOS of the FiM state calculated at LSDA+$U_{\rm eff}$ = 0.5 eV. (c) DOS of the FiM state of Mn$_3$Si$_2$Se$_6$ calculated at LSDA+$U_{\rm eff}$ = 0.5 eV, based on the same crystal lattice constants of Mn$_3$Si$_2$Te$_6$. (b-c) Both the total DOS and atomic projected DOS are represented by different colors.}
\label{dos-FiM}
\end{figure}

In Mn$_3$Si$_2$Te$_6$, the Mn-ions have a $2+$ valence, leading to a $d^5$ electronic configuration
with five half-filled $3d$ orbitals.
Without any interactions, the Mn $3d$ states display strong itinerant behavior hybridized with the Te $5p$ orbitals. By introducing the Hubbard interaction $U$, the five half-filled Mn-$3d$ orbitals should be Mott-localized with a small bandwidth, opening a Mott gap.  Furthermore, the Te $5p$ states are usually extended in real space~\cite{Zhang:prb20-2}, leading to wide bands with large bandwidth. For these reasons, Mn$_3$Si$_2$Te$_6$ is more likely a charge-transfer system, with the expected local DOS schematically shown in Fig.~\ref{dos-FiM}(a).  Then, the band gap of this system depends on the energy gap between empty Mn's $3d$ and fully occupied Te's $5p$ state, as shown in Fig.~\ref{dos-FiM}(a).

Next, we also calculated the DOS of the FiM state of Mn$_3$Si$_2$Te$_6$ using LSDA+$U_{\rm eff}$ ($U_{\rm eff} = 0.5$ eV), as displayed in Fig.~\ref{dos-FiM}(b). Note that the spin dependence of the correlation energy density is already considered in the LSDA portion. Hence, the additional effective $U_{\rm eff}$ of the DFT calculations is different from the Hubbard $U$ in the standard model Hamiltonian calculations: the value of $U_{\rm eff}$ of DFT calculations is always smaller than the value of $U$ of the Hubbard Hamiltonian model.

According to the DOS, the states near the Fermi level are mainly contributed by the Te $5p$ orbitals, partially hybridized with the Mn-$3d$ orbitals. The Te $5p$ orbitals display a strong extended behavior with a large bandwidth. Furthermore, the Mn's $3d$ states display a Mott-localized behavior with a large Mott gap, where those occupied $3d$ states are mainly located at lower energy regions from $-4$ to $-3$ eV [see Fig.~\ref{dos-FiM}(b)]. Then, the band gap of this system dramatically decreases to $\sim 0.16$ eV, where this gap is caused by the occupied Te $5p$ and unoccupied Mn $3d$ states. In this case, this general physical picture could intuitively be used to understand the semiconducting behavior with a small band gap in the experiment~\cite{Seo:nature} although the Mn $3d$ orbitals form a large Mott gap. In addition, the Mn $3d$ states still display Mott-localized behavior with increasing Mott gap as the values of $U_{\rm eff}$ increases [See Fig. S1(b)].

To better understand this physical picture, we also calculated the DOS of the FiM state of Mn$_3$Si$_2$Se$_6$ based on the same crystal lattice constants of Mn$_3$Si$_2$Te$_6$ using LSDA+ $U_{\rm eff}$ ($U_{\rm eff} = 0.5$ eV), as displayed in Fig.~\ref{dos-FiM}(c). Compared with Mn$_3$Si$_2$Te$_6$, the Se's $4p$, and Mn's $3d$ orbitals are more localized, leading to a smaller bandwidth than for the Te case. Then, using the same parameters ($U_{\rm eff} = 0.5$ eV) and crystal structure lattice constants, the calculated band gap of Mn$_3$Si$_2$Se$_6$ increases to about $0.83$~eV. Hence, all the results we obtained for the DOS support the charge-transfer picture, as already analyzed in previous paragraphs.

By introducing the SOC effect, the Te $5p$ bands would split, leading to different physical properties for different spin orientations, as will be discussed in the next section.

\subsection{B. Spin orientation}
Turning on the SOC, we found that the spin quantization axis points to the $ab$ plane but with only a very small difference in energy compared to the [001] direction, indicating that the spin prefers to be in the $ab$ plane. In addition, the MAE (E[110]-E[001]) is calculated to be $-0.71$ meV/Mn for the FiM state, by comparing the energy difference between [110] and [001]. This small energy difference arises from the small SOC of the Mn $3d^5$ ions. As discussed in the previous section, the calculated orbital moments of Mn are quite small: $0.031$ $\mu_{\rm B}$/Mn1 and $0.040$ $\mu_{\rm B}$/Mn2, respectively. Then, this small orbital moment contributes to the small difference in the anisotropy energy ($\sim 0.71$ meV). In principle, an external magnetic field could rotate the spin from the $ab$ plane to the [001] direction, and then flip the AFM spins to FM order if the energy differences between those states are small.

\begin{figure}
\centering
\includegraphics[width=0.48\textwidth]{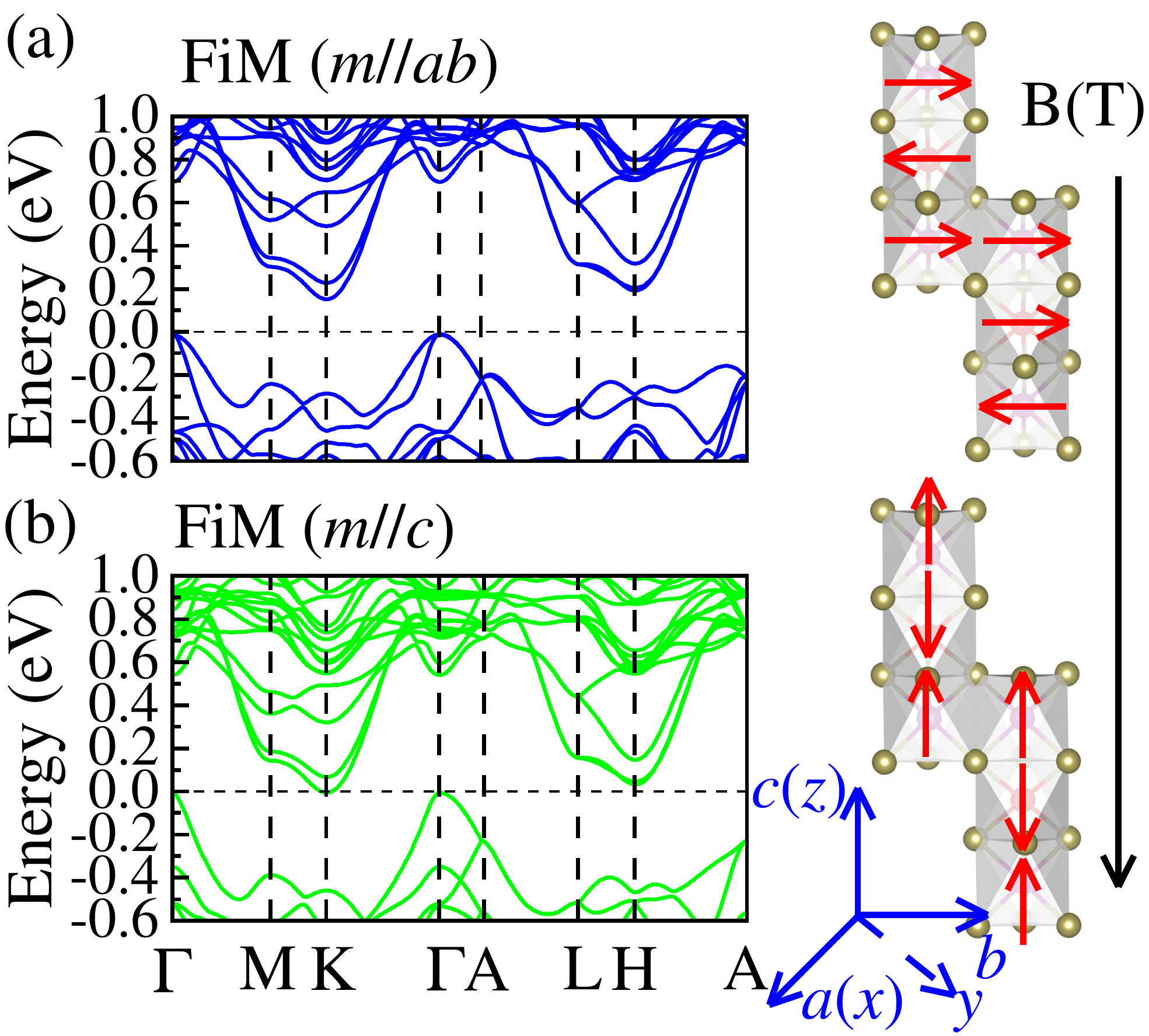}
\caption{Band structure of Mn$_3$Si$_2$Te$_6$ near the Fermi level, based on the LSDA+$U_{\rm eff}$+SOC ($U_{\rm eff} = 0.5$ eV) for (a) the FiM ($m$// $ab$) state (spin lying along the [110] direction), and (b) the FiM ($m$// $c$) state (spin lying along the $c$-axis). The Fermi level is shown with dashed horizontal lines. The coordinates of the high-symmetry points in the bulk Brillouin zone (BZ) are $\Gamma$ = (0, 0, 0), M = (0.5, 0, 0), K = (1/3, 1/3, 0), A = (0, 0, 0.5), L = (0.5, 0, 0), and H = (1/3, 1/3, 0.5). The evolution of the magnetic state under the magnetic field along the $c$-axis is show at the right.}
\label{Bands-switch}
\end{figure}

In the well-known CMR materials, such as the hole-doped manganites La$_{1-x}$Ca$_x$MnO$_3$, the phase separation mechanism plays a key role to understand the CMR effect, which is caused by the competition between the AFM insulating and FM metallic phases, both induced by the double exchange interaction~\cite{Dagotto:rp,Salamon:rmp}. However, in Mn$_3$Si$_2$Te$_6$, the FM metallic state has much higher energy than the AFM insulating state (see Table~\ref{Table1}), indicating that via magnetic external fields it will be difficult to flip AFM spins to a FM order. In this case, the origin of the CMR effect of Mn$_3$Si$_2$Te$_6$ should be different from that of typical hole-doped manganites.

Introducing the SOC effect, the Te-occupied $5p$ bands start to split with different nodal-line degeneracy in different spin orientations of the Mn spins, as displayed in Fig.~\ref{Bands-switch}. Note that the topological nodal-line degeneracy physics of this compound has been studied in detail in a recent publication~\cite{Seo:nature}. The FiM [110] state, with spin moments lying in the $ab$-plane, displays a semiconducting behavior with an indirect gap $\sim 161.2$ meV [see Fig.~\ref{Bands-switch} (a)]. For the FiM [001] state with the spin oriented along the $c$-axis, the band structure indicates a metallic phase, as shown in Fig.~\ref{Bands-switch}(b). Hence, the insulator-metal transition occurs by switching the spin orientation from the [110] plane to the [001] direction. Due to the small energy difference between [110] and [001] directions ($\sim 0.71$ eV), this insulator-metal transition could be induced by an external magnetic field along the $c$-axis.

This analysis suggests that the FiM [110] insulating and FiM [001] metallic phases could compete due to the small energy difference scale, which could qualitatively explain the observed CMR effect in experiments~\cite{Liu:prb21,Ni:prb21,Seo:nature,Wang:prb22}. Hence, the origin of the CMR effect of Mn$_3$Si$_2$Te$_6$ should be related to the competition between insulating and metallic phases in the same magnetic state but different spin orientations. This behavior is similar to the observed CMR in EuCd$_2$As$_2$, where the competition among phases is also induced by the spin orientations with competing energies between magnetic topological insulating, trivial insulating, and Weyl semimetal phases~\cite{Du:qm}.

\begin{figure}
\centering
\includegraphics[width=0.48\textwidth]{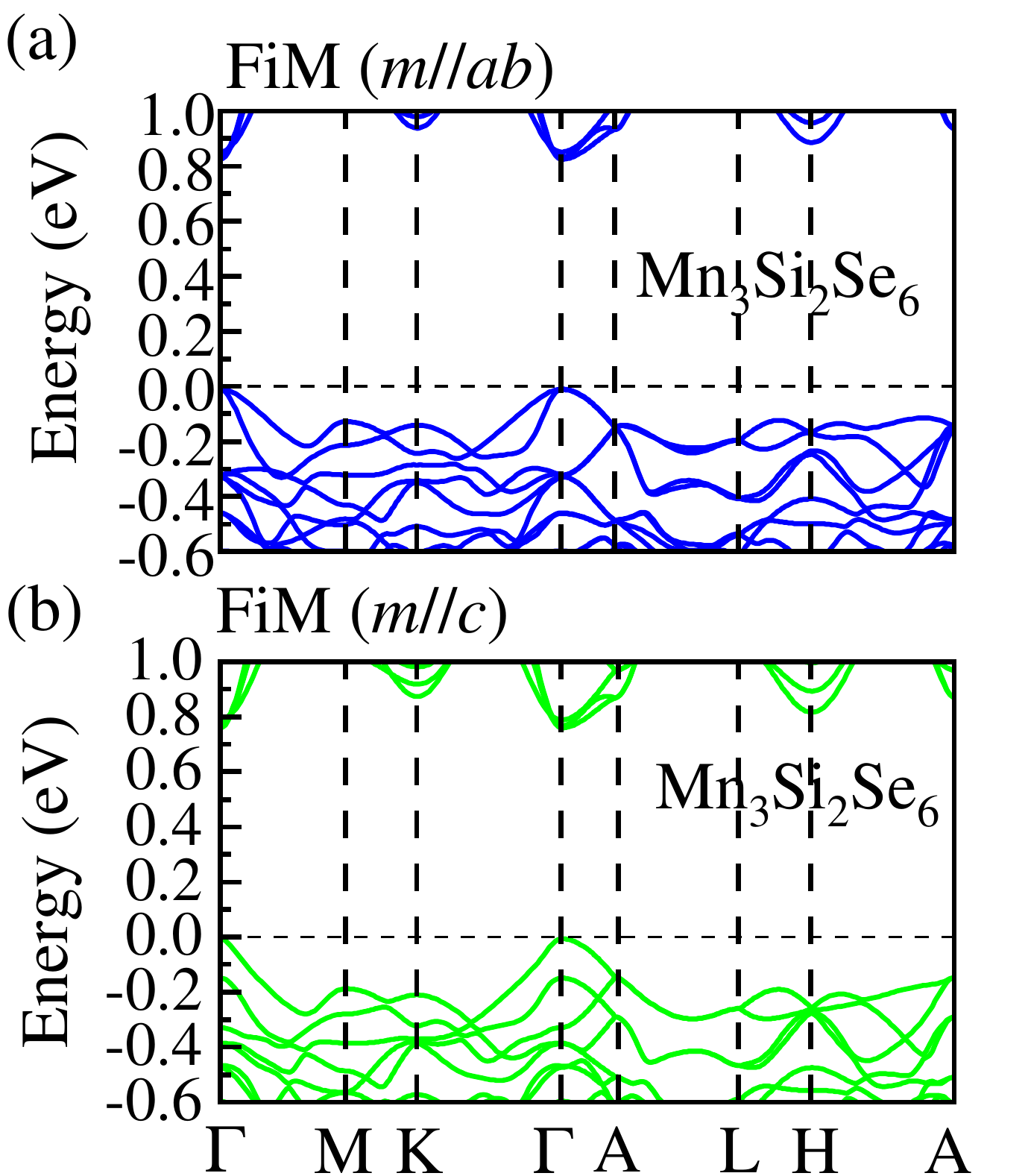}
\caption{Band structure of Mn$_3$Si$_2$Se$_6$ near the Fermi level based on the same crystal structure of Mn$_3$Si$_2$Te$_6$ using LSDA+SOC+$U_{\rm eff} = 0.5$ eV for (a) the FiM (// $ab$) state (spin lying along [110] direction), and (b) the FiM (// $c$) state (spin lying along the $c$-axis). The Fermi level is shown with dashed horizontal lines.}
\label{Bands-switch-Se}
\end{figure}

Furthermore, this insulator-metal transition with a reduced band gap induced by the spin orientations ($161.2$ to $0$ meV) is caused by the nodal-line structure of Te's $p$-bands, also related to the SOC effect.  If the SOC strength is reduced, the change in the $\Delta$ ([110]-[001]) band gap between in-plane and out-plane directions should be reduced as well. Hence, we also calculated the band structures of the FiM state of Mn$_3$Si$_2$Se$_6$ for different spin orientations based on the same crystal structure of Mn$_3$Si$_2$Te$_6$, because the SOC effect of Se is smaller than for Te atoms. As displayed in Fig.~\ref{Bands-switch-Se}, the change in the $\Delta$ band gap between the [110] and [001] directions is reduced to about $68.2$ meV, as expected. The CMR and colossal angular MR effects of this system are related to the change of bands gaps of different spin orientations,
where the reduced value of the band gap from in-plane to out-plane directions is decided by the value of the SOC. Hence, the CMR or colossal angular MR effects should be strongly suppressed in Mn$_3$Si$_2$Se$_6$.

\begin{figure}
\centering
\includegraphics[width=0.48\textwidth]{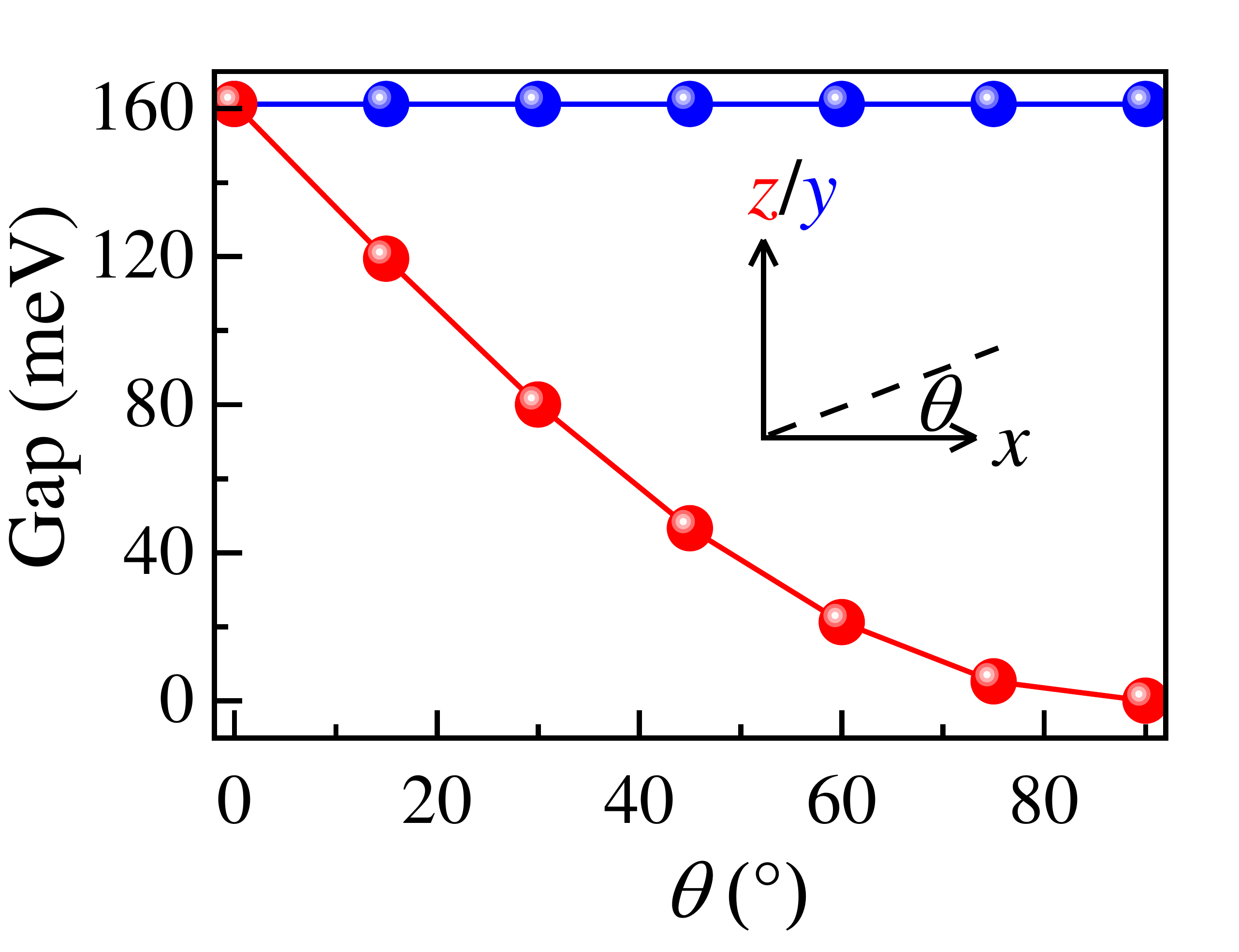}
\caption{ Evolution of the calculated band gap of the ferrimagnetic state of Mn$_3$Si$_2$Te$_6$, as a function of angle $\theta$.}
\label{switch}
\end{figure}

To better understand the spin orientation effet, we simulated the switching ``path'' in the $ab$ and out of plane by changing the angle $\theta$ between the $x-$ and $y-/z-$ axes. Changing the angle $\theta$ (corresponding to different spin quantization axis) in the $xy$ plane (corresponding to the $ab$ crystal plane), the energy and the band gap are identical, independent of the angle $\theta$ [See Fig. S2 and Fig.~\ref{switch}]. In addition, the band gap rapidly is reduced by changing the spin quantization axis from the $x-$ to the $z$-axis, as displayed in Fig.~\ref{switch}, consistent with the large resistance reduction as the magnetic field orientation $\theta$ is varied~\cite{Seo:nature}. Note that the insulator-metal transition of spin orientation is a ``gradual'' transition as spin angle $\theta$ changes, not a "sharp" phase transition. This insulator-metal transition caused by the angle $\theta$ could also explain the
colossal angular MR observed in experiments~\cite{Seo:nature}.

\subsection{C. Ferrimagnetic state and magnetic phase diagram}
Based on the optimized crystal lattice of the FiM state, we mapped the DFT energy into a classical model to obtain three magnetic exchange couplings $J's$, assuming the magnitude of the spins is considered as $1$ for simplicity. Note that those values of $J$ are calculated by mapping the DFT energies of four different states (FiM, AFM1, AFM2, and FM) to the classic spin model.

At $U_{\rm eff} = 0.5$ eV, we obtained that the three magnetic couplings are $J_1 = -33.8$, $J_2 = -11.4$ and $J_3 = -13.4$, all in units of meV. The coupling $J_1$ involving nearest-neighbor Mn-Mn distances provides the strongest AFM magnetic exchange coupling, much larger than $J_2$ and $J_3$. In addition, the $J_2$ between NNN Mn-Mn sites is also AFM but smaller than the AFM $J_3$ with NNNN Mn-Mn distance, in agreement with previous calculations~\cite{May:prb17,Seo:nature}. In this case, those three AFM couplings could lead to strong frustration due to the competition in a triangular geometry. Note that our qualitative results are approximately independent of the choice of $U_{\rm eff}$. For the benefit of the readers, based on the optimized crystal lattice of the FiM state, we also evaluated the three magnetic exchange coupling $J's$ vs. $U_{\rm eff}$, as summarized in Table~\ref{Table2}.

\begin{table}
\centering\caption{Calculated magnetic couplings: $J_1$, $J_2$, $J_3$ (in meV), and the ratio ($J_2$/$J_1$ and $J_3$/$J_1$) at several values of $U_{\rm eff}$.}
\begin{tabular*}{0.49\textwidth}{@{\extracolsep{\fill}}lllllc}
\hline
\hline
$U_{\rm eff}$    & $J_1$ & $J_2$ & $J_3 $ & $J_2$/$J_1$ & $J_3$/$J_1$\\
\hline
0  & -38.6  & -12.4 & -15   & 0.321 & 0.389  \\
0.5 & -33.8	& -11.4	& -13.4 & 0.337	& 0.396 \\
1  & -27.7	& -9.1 &	-10.9 & 0.329 &	0.394\\
1.5 & -22.7	& -7.1	& -8.7	& 0.313 & 0.383 \\
2   & -18.3	& -5.6 &	-6.8 &0.306	& 0.372 \\
3  & -12.6	& -3.2	& -3.8	&  0.254 & 0.302\\
\hline
\hline
\end{tabular*}
\label{Table2}
\end{table}

\begin{figure}
\centering
\includegraphics[width=0.48\textwidth]{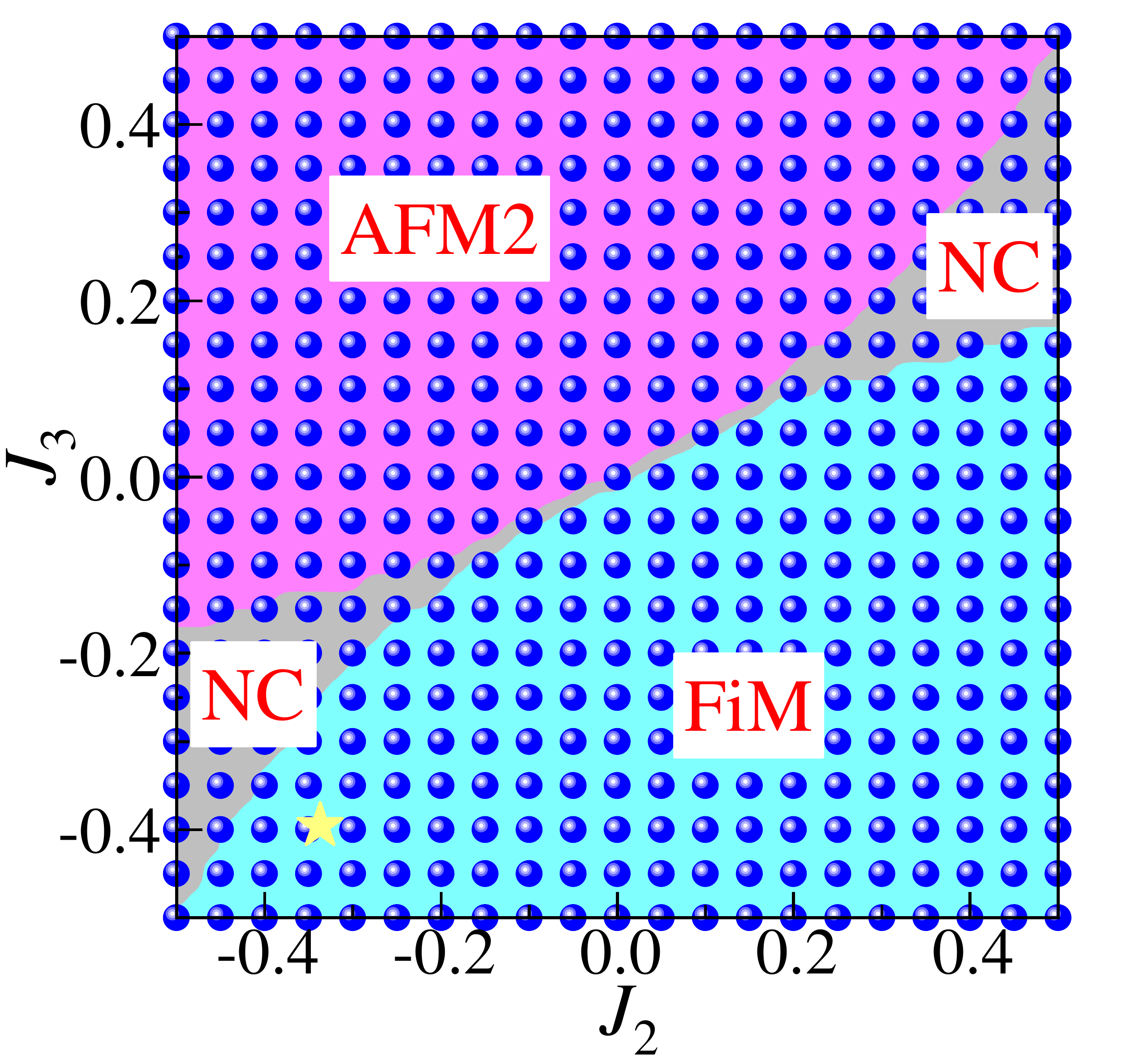}
\caption{Monte Carlo magnetic phase diagram of the classical $XY$ spin model varying normalized $J_1$, $J_2$ and $J_3$, using a $40\times40$ lattice. Different phases are indicated with the labels AFM2, NC, and FiM phases. Here, we fixed the NN as AFM coupling with $J_1 = -1$ and changed $J_2$ or $J_3$ from -0.5 to 0.5. Note that the AFM2, NC, and FiM patterns have the same energies at $J_2$ = $J_3$ = 0. Small solid blue circles indicate specific values of data points that
were explicitly investigated in our MC calculations. The yellow star shows the normalized values of the calculated $Js$ in DFT for the real material Mn$_3$Si$_2$Te$_6$ ($J_1 = -1$, $J_2 = -0.337$, and $J_3 = -0.396$).}
\label{Phasediagram}
\end{figure}

Next, we calculated the magnetic phase diagram varying $Js$, based on the classical $XY$ spin model using MC techniques (spin patterns are provided in real space).
For the $J_1$ path (NN Mn-Mn sites) along the $c$-axis, the magnetic coupling should be the strongest AFM due to the strong overlap of $d_{3z^2-r^2}$ orbitals, where the local $\{x,y,z\}$ basis is considered, as shown in Fig.~\ref{Crystal}. Hence, we fixed the NN $J_1$ path to be AFM with $J_1 = -1$ and the NNN $J_2$ and NNNN $J_3$ were both considered to be either AFM or FM, by changing the values from -0.50 to 0.50. We found two dominant phases, involving collinear AFM2 and FiM spin orders, as shown in Fig.~\ref{Phasediagram}. In addition, a noncollinear (NC) spin pattern, to be shown explicitly below, was also obtained at the boundaries between the AFM2 and FiM phases (see Fig.~\ref{Phasediagram}). It should be noted that the boundaries between different phases should be considered only as crude approximations. However, the existence of the three different phase regions was clearly established, even if the boundaries are only estimations. We believe our theoretical magnetic phase diagram should encourage a more detailed experimental study of this compounds or related systems, for example by varying the chemical composition.

\begin{figure*}
\centering
\includegraphics[width=0.96\textwidth]{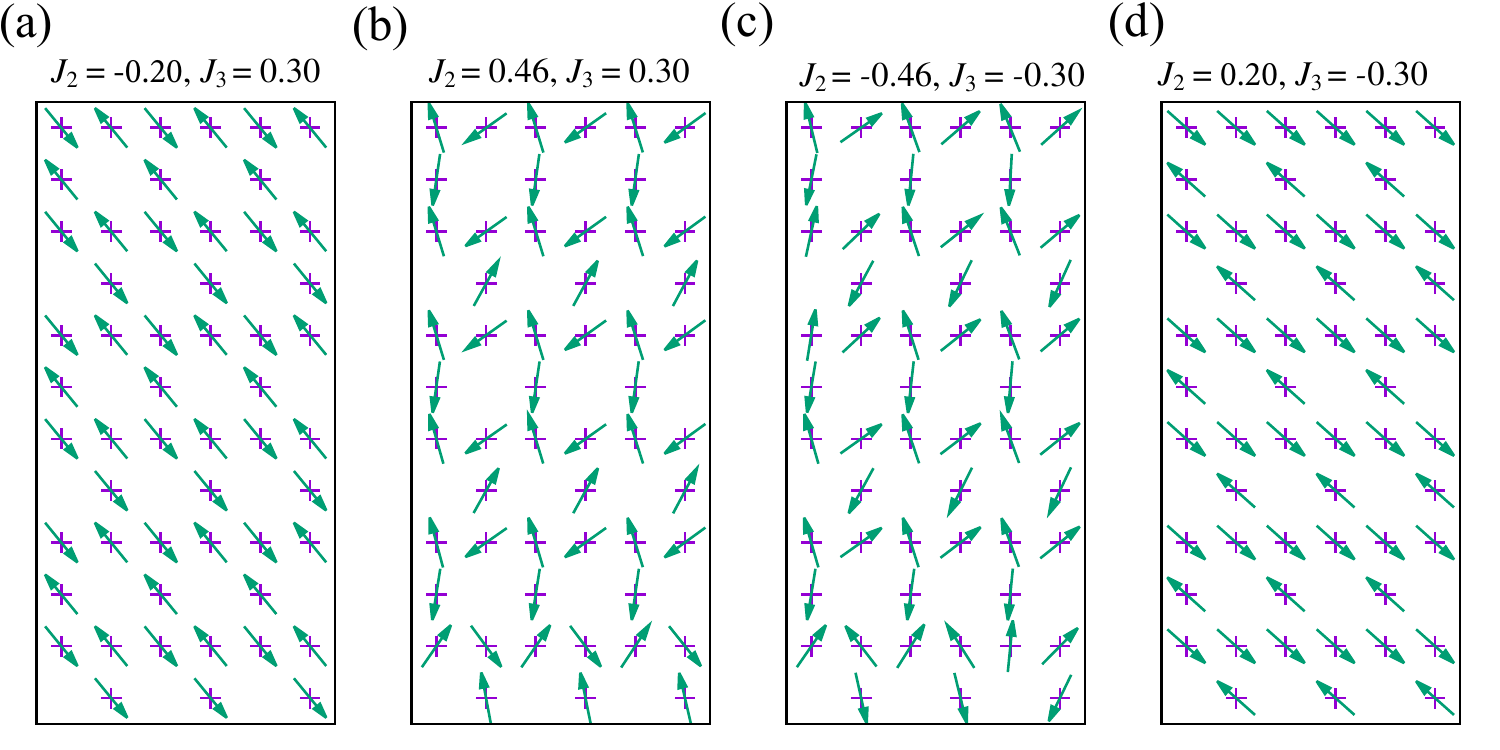}
\caption{Spin patterns for different values of normalized $Js$, where $J_1$ is fixed to be -1. (a) AFM2 state at $J_2$ = -0.20 and $J_3$ = 0.30. (b) NC state at $J_2$ = 0.46 and $J_3$ = 0.30. (c) NC state at $J_2$ = -0.46 and $J_3$ = -0.30. (d) FiM state at $J_2$ = 0.20 and $J_3$ = -0.30.}
\label{patterns}
\end{figure*}

In the unit cell of the Mn$_3$Si$_2$Te$_6$ lattice, there are 12 and 6 spin pairs along the $J_2$ or $J_3$ paths, respectively. Hence, the region of stability of the different magnetic phases is mainly decided by the sign of $J_3$. If $J_3 \textgreater 0$ (FM interaction), the AFM2 phase is the dominant phase in our MC phase diagram. For example, the real-space spin pattern at $J_3 = 0.30$ and $J_2 = -0.20$ clearly displays antiferromagnetic order, as shown in Fig.~\ref{patterns} (a) (obtained from low-$T$ MC plus optimization). Furthermore, the NC spin order was obtained in some regions due to the strong competition between $J_2$ and $J_3$ if the $J_2$ also is a FM coupling and larger than $J_3$ [see, as an example, the real-space spin pattern at $J_3$ = 0.30 and $J_2$ = 0.46 in Fig.~\ref{patterns}(b)].

In addition, a similar NC order was also obtained if the AFM $J_2$ could compete with AFM $J_3$, such as at
$J_2$ = -0.46 and $J_3$ = -0.30, with the NC pattern displayed in real space in Fig.~\ref{patterns}(c). In the $J_3 \textless 0$ and $J_2 \textgreater 0$ regions (AFM $J_3$ and FM $J_2$), the FiM phase is stable, as displayed is Fig.~\ref{Phasediagram}. The example $J_2$ = 0.20 and $J_3$ = -0.30 in Fig.~\ref{patterns}(d) clearly shows that the real-space spin pattern corresponds to FiM order. Considering the calculated magnetic coupling values ($J_1 = -33.8$, $J_2 = -11.4$ and $J_3 = -13.4$ meV)~\cite{Liechtensteincontext}, the real material Mn$_3$Si$_2$Te$_6$ is located inside the FiM region in our MC phase diagram (see yellow star in Fig.~\ref{Phasediagram}), in agreement with experiments.

\subsection{D. Doping effect}
To better understand the physical properties under Se or Ge doping in Mn$_3$Si$_2$Te$_6$, we employed the virtual crystal approximation (VCA) to simulate the doping effect, technique widely used in the electronic structure context~\cite{Bellaiche:Prb,Ramer:Prb,Zhang:prb21}. Here, both the lattice constants and atomic positions were fully relaxed with different spin states, for different Se or Ge doping levels.

\begin{figure}
\centering
\includegraphics[width=0.48\textwidth]{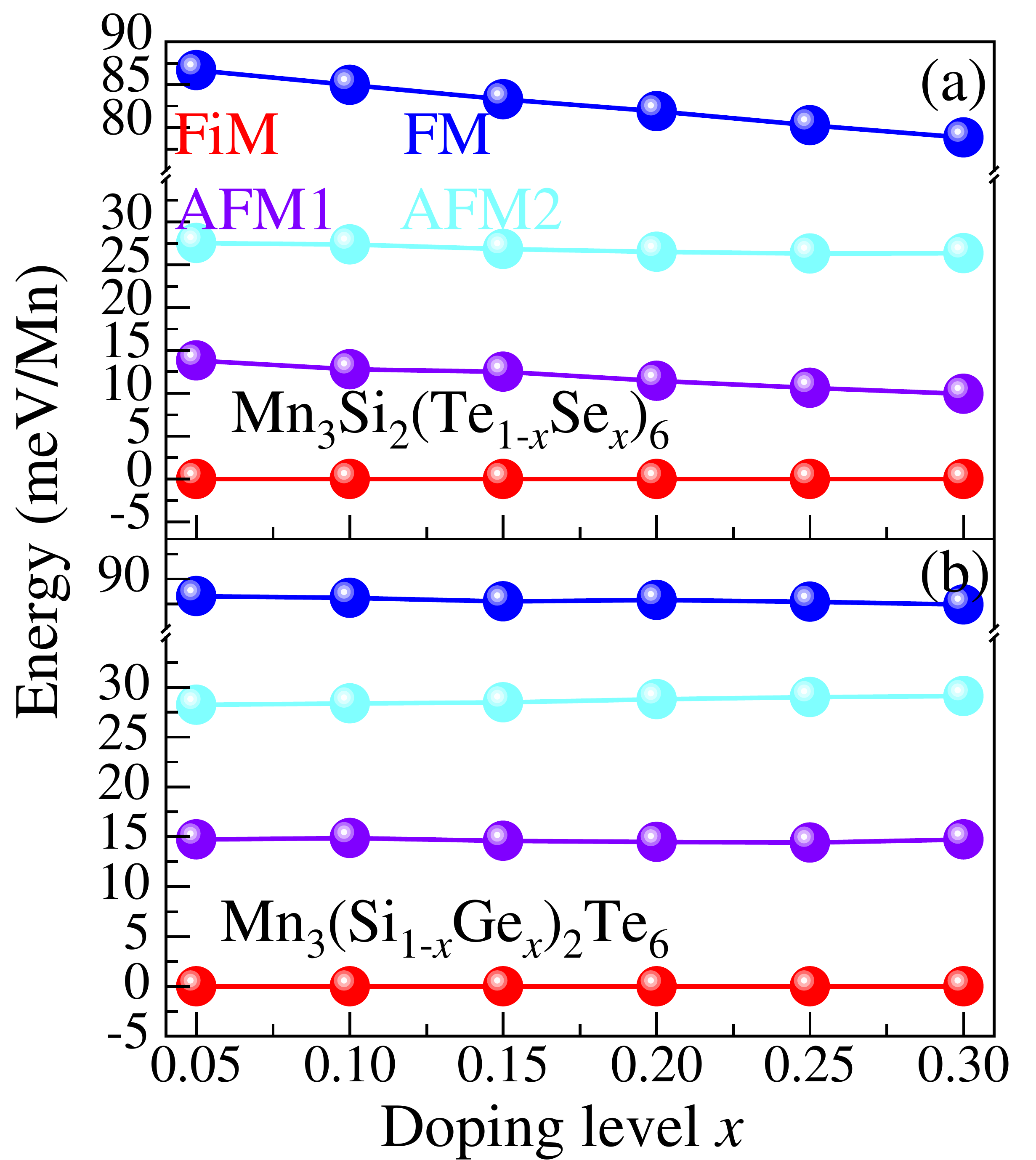}
\caption{Energies for different magnetic states under different doping levels: (a) The case of Se doped in the Te site; (b) Ge doped in the Si site. The FiM state is chosen as energy reference.}
\label{doping-E}
\end{figure}

For both Mn$_3$Si$_2$(Te$_{1-x}$Se$_x$)$_6$ and Mn$_3$(Si$_{1-x}$Ge$_x$)$_2$Te$_6$, the FiM state has the lowest energy among all magnetic candidates at the doping levels we studied, as displayed in Fig.~\ref{doping-E}. In addition, the energy differences of different magnetic states do not change much. Then, we conclude that the magnetic transition temperatures at small doping of Se or Ge would not change much either, in agreement with experimental results for the magnetic susceptibility of the undoped, $20\%$ Se-doped, and $6\%$ Ge-doped cases~\cite{Seo:nature}. Furthermore, the spin quantization axis always points along the $ab$ crystal plane but with only a small difference in energy with respect to the $c$-axis ([001] direction), indicating that the spin still favors lying in the $ab$ crystal plane, for both the Ge- and Se-doped cases.

\begin{figure}
\centering
\includegraphics[width=0.48\textwidth]{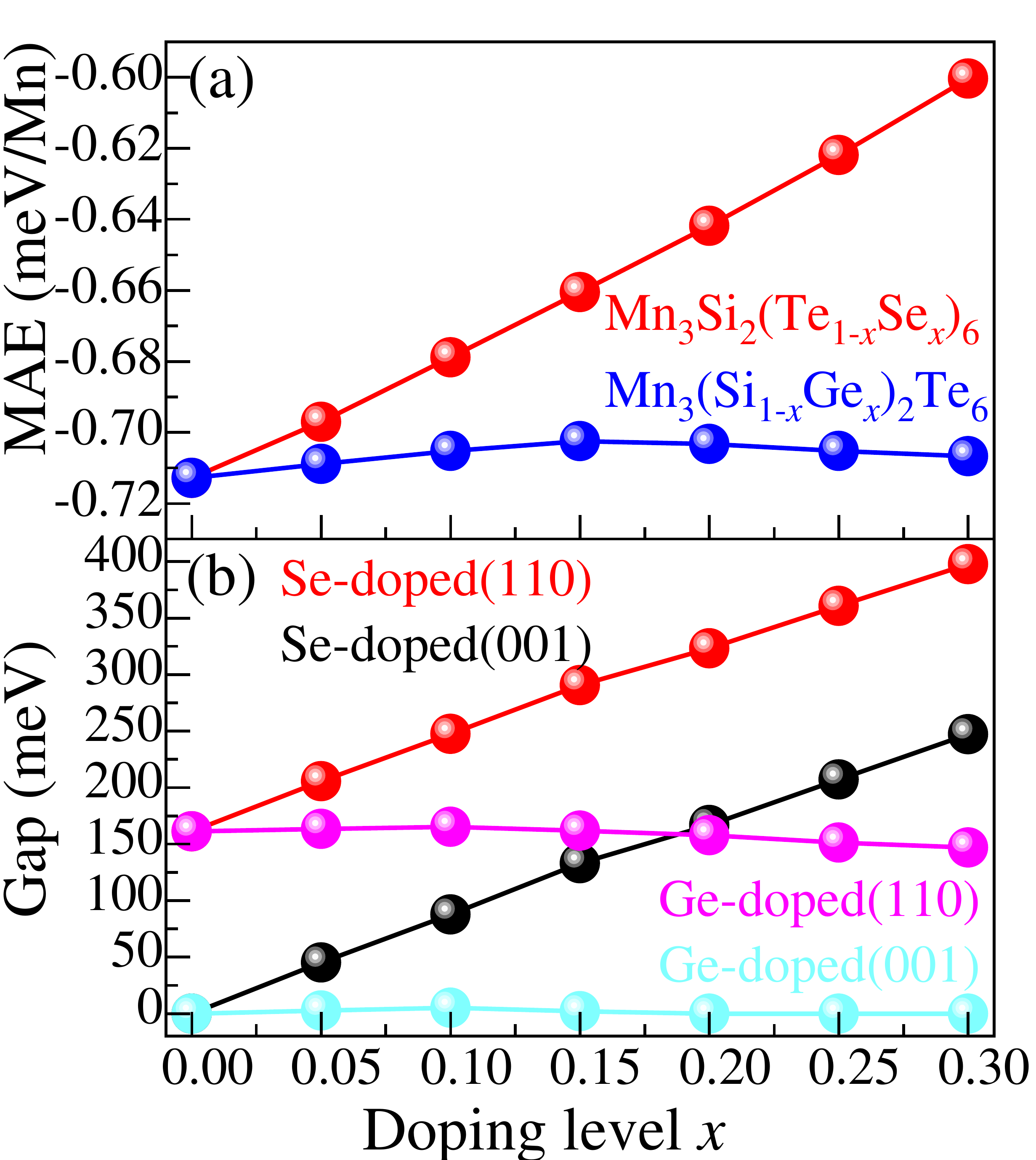}
\caption{(a) The MAE (E[110]-E[001]) for the FiM state under doping. (b) The calculated band gap of the FiM state under Se- or Ge-doping, for the spins lying along the [110] and [001]
directions, respectively.}
\label{doping-Gap}
\end{figure}

As the doping level increases, the MAE (E[110]-E[001]) slightly decreases from $-0.71$ meV/Mn ($x = 0$) to $-0.60$ meV/Mn ($x = 0.3$) in the Se-doped case, while it is almost unchanged in the Ge-doped case, as displayed in Fig.~\ref{doping-Gap}(a). This is reasonable. As the doping $x$ increases in Mn$_3$Si$_2$(Te$_{1-x}$Se$_x$)$_6$, the orbital moments of Mn decrease to $0.026/0.034$ $\mu_{\rm B}$/Mn for the Mn1 and Mn2 sites, respectively. Because the anisotropy is caused by the small orbital moment of the Mn atoms, the MAE is slightly reduced in the Se-doped case. However, for the Ge-doped case, the calculated orbital moments of Mn are almost unchanged as the doping $x$ increases. Hence, the MAE of Ge-doped case does not change in the doping range we studied.

In addition, as the doping $x$ increases, the calculated band gap of the FiM state of Mn$_3$Si$_2$(Te$_{1-x}$Se$_x$)$_6$ smoothly increases for both the spin orientations [110] and [001] [see Fig.~\ref{doping-Gap}(b)]. For Mn$_3$(Si$_{1-x}$Ge$_x$)$_2$Te$_6$, the calculated band gaps just slightly change for the two different spin orientations, at the doping levels we studied. Note that the different behavior of the calculated band gaps for the Se-doped and Ge-doped cases are also independent of $U_{\rm eff}$ [See Fig. S5(b)]. As discussed in Section III.A, the band gap of this system is mainly caused by the occupied Te $5p$ and unoccupied Mn $3d$ orbitals. By doping Se into the Te sites, the states near the Fermi level would be more localized, leading to a reduced bandwidth, resulting in an increased band gap, as the doping $x$ increases. However, the $p$ and $d$ states should not be seriously affected by doping Ge in the Si sites because most Si states are located at deep energies far from the Fermi level. Hence, the calculated band gap does not change much in the small doping region.

\begin{figure}
\centering
\includegraphics[width=0.48\textwidth]{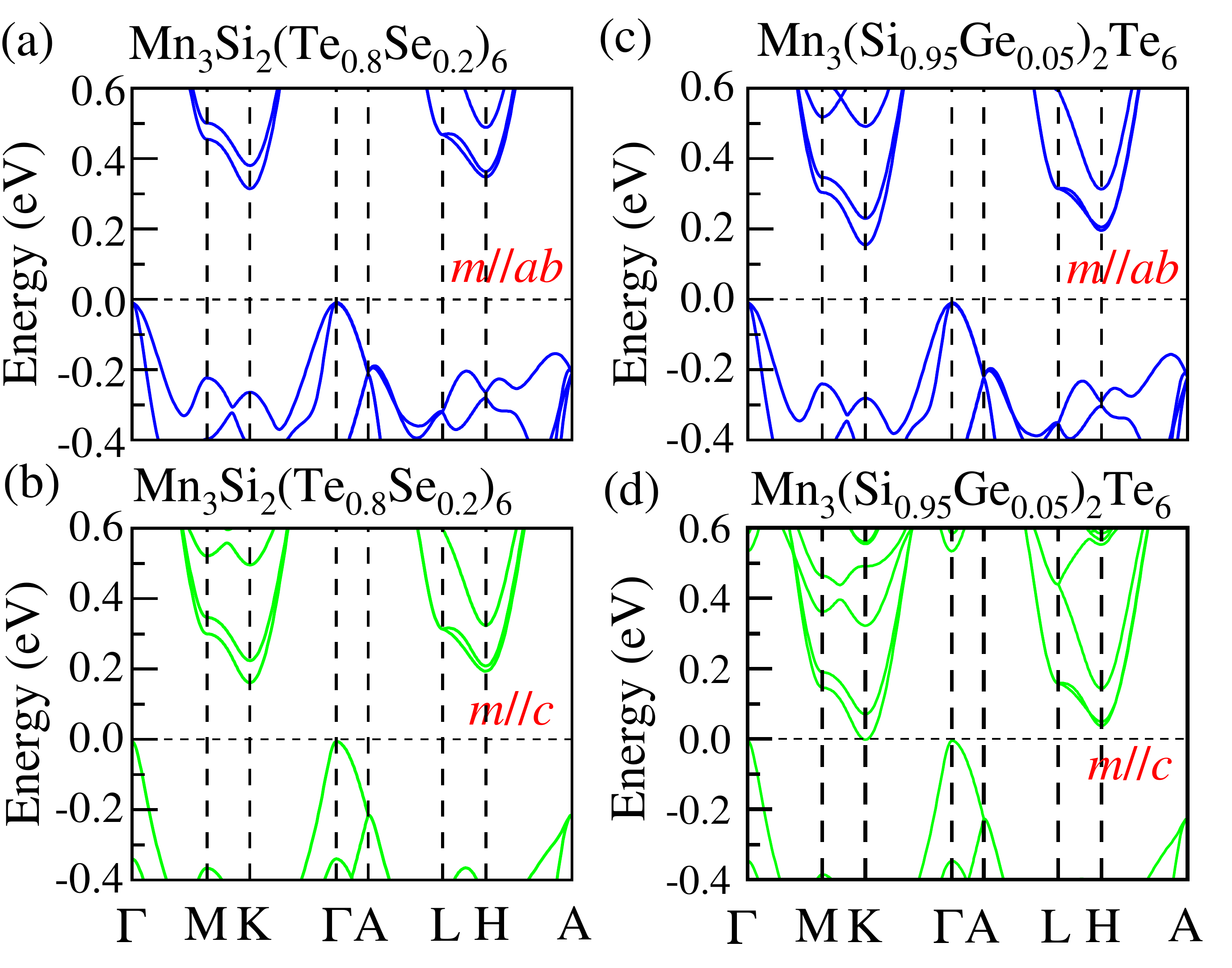}
\caption{Band structure of the FiM state near the Fermi level based on LSDA+$U$+SOC ($U_{\rm eff} = 0.5$ eV calculations,
for Mn$_3$Si$_2$(Te$_{\rm 0.8}$Se$_{\rm 0.2}$)$_6$ and Mn$_3$(Si$_{\rm 0.95}$Ge$_{\rm 0.05}$)$_2$Te$_6$.
(a) The FiM ($m$//$ab$) state (spin lying along [110] direction) for Mn$_3$Si$_2$(Te$_{\rm 0.8}$Se$_{\rm 0.2}$)$_6$. (b) The FiM ($m$//$c$) state (spin lying along $c$-axis) for Mn$_3$Si$_2$(Te$_{\rm 0.8}$Se$_{\rm 0.2}$)$_6$.
(c) The FiM ($m$//$ab$) state (spin lying along [110] direction) for Mn$_3$(Si$_{\rm 0.95}$Ge$_{\rm 0.05}$)$_2$Te$_6$. (d) The FiM ($m$//$c$) state (spin lying along $c$-axis) for Mn$_3$(Si$_{\rm 0.95}$Ge$_{\rm 0.05}$)$_2$Te$_6$. The Fermi level is shown with dashed horizontal lines. The coordinates of the high-symmetry points in the bulk BZ are $\Gamma$ = (0, 0, 0), M = (0.5, 0, 0), K = (1/3, 1/3, 0), A = (0, 0, 0.5), L = (0.5, 0, 0), and H = (1/3, 1/3, 0.5).}
\label{Bands-doping}
\end{figure}

Furthermore, in the FiM state with spin orientation both along the $ab$ plane and $c$-axis, the system displays insulating behavior with an indirect band-gap characteristic of the Se-doped case. Here, we also calculated the band structure of the FiM state for the $20\%$ Se doping case, for both spins along $ab$ and $c$ directions, as shown in Figs.~\ref{Bands-doping}(a-b). The calculated indirect band gaps of the FiM state of the $20\%$ Se doping case are about $323.3$ and $166.9$ meV for the spin lying along $ab$ or along $c$, respectively. In this case, the insulator-metal transition disappears by switching the angle $\theta$ between the [110] and [001] axis. This large gap in the FiM ($m//ab$) state would greatly reduce the conductivity of the system, leading to a far more reduced colossal angular MR effect. This could qualitatively explain the strongly reduced colossal angular MR of the $20\%$  Sr-doping case observed at $2$ K~\cite{Seo:nature}.

Moreover, the difference $\Delta$ of band gaps between the two different spin orientations gradually decreases to  $156.4$ meV in Mn$_3$Si$_2$(Te$_{\rm 0.8}$Se$_{\rm 0.2}$)$_6$, due to the reduced SOC effect of Se. For the Ge-doping case, the calculated band gap of the FiM state slightly increases, reaching the maximum value at $x = 0.1$ ($165.4$ and $5.2$ meV for the $m$//$ab$ or $m$//$c$, respectively), and then decreases for the spin both lying along the $ab$ plane and $c$-axis, as the doping $x$ increases. At $x = 0.05$ in the Ge-doped case, the calculated indirect band gaps are $163.4$ and $2.8$ meV for the spins along $ab$ plane and $c$-axis, respectively [see Figs.~\ref{Bands-doping}(c-d)]. This small gap of Mn$_3$(Si$_{\rm 0.95}$Ge$_{\rm 0.05}$)$_2$Te$_6$ in the FiM ($m$//$c$) state would reduce the conductivity of the system, leading to a reduced colossal angular MR effect, compared to the undoped case. Different from the large gap of the FiM ($m$//$c$) state in the $20\%$ Sr-doping case, the reduced amplitude of the colossal angular MR effect is not too large in the $6\%$ Ge-doped case. This also could qualitatively explain the slightly reduced colossal angular MR effect in the $6\%$ Ge-doped case, at very low temperatures~\cite{Seo:nature}.

\section{III. Conclusions}
In this publication, we systematically studied the layered system Mn$_3$Si$_2$Te$_6$ with alternating stacking of honeycomb and triangular layers, by combining first-principles DFT and classical MC calculations. Based on the {\it ab initio} DFT results, we found that the ferrimagnetic state is the most likely magnetic ground state, in agreement with previous neutron results. In addition, the states near the Fermi level are primarily contributed by the Te $5p$-states hybridized with the Mn $3d$-orbitals, leading to a charge transfer system. Furthermore, the spin orientations of the FiM state display different behaviors: insulating state in the $ab$ plane and metallic state in the out-of-plane direction, while the energy difference between them is only about
$\sim 0.71$ meV. In this case, the very similar energies between the FiM [110] insulating and FiM [001] metallic phases are likely responsible for the observed CMR effect in experiments. By changing the angle $\theta$ of the spins orientation, the calculated band gap rapidly is reduced, leading to an insulator-metal transition, which could also explain the observed colossal angular MR effect.

By mapping the DFT energy to a Heisenberg model, we obtain three magnetic exchange couplings, all of them AFM. In addition, we also constructed the magnetic phase diagram varying $J_2$/$J_1$ and $J_3$/$J_1$ ($J_1$ was fixed to be -1), based on the classical Heisenberg model using the MC method, where three magnetic phases were obtained. Moreover, we also investigated the Se- or Ge-doping
in Mn$_3$Si$_2$Te$_6$. The FiM state has the lowest energy among the magnetic candidates for both cases. Due to the reduced orbital moment of Mn in the Se-doped case, the MAE slightly decreases as the doping $x$ level increases. However, for the Ge-doped case, the calculated orbital moments of Mn remain almost unchanged, as the doping level $x$ increases. Hence, the MAE of the Ge-doped case does not change in the doping level range studied here. Furthermore, the insulator-metal transition caused by the spin orientation disappears in the Se-doped case because of the strongly reduced spin-orbital coupling effect of Se, resulting in an insulating phase in the FiM [001] phase, leading to a reduced colossal angular MR. However, the band gap of the Ge-doped case does not change much for both the [110] and [001] directions, as the doping level $x$ increases. Thus, we believe our results for Mn$_3$Si$_2$Te$_6$ provide guidance to experimentalists and theorists working
in this system or related materials.

\section{Acknowledgments}
The work of Y.Z., L.-F.L., A.M., and E.D. was supported by the U.S. Department of Energy, Office of Science, Basic Energy Sciences, Materials Sciences and Engineering Division.

\section{APPENDIX}
\begin{figure}
\centering
\includegraphics[width=0.48\textwidth]{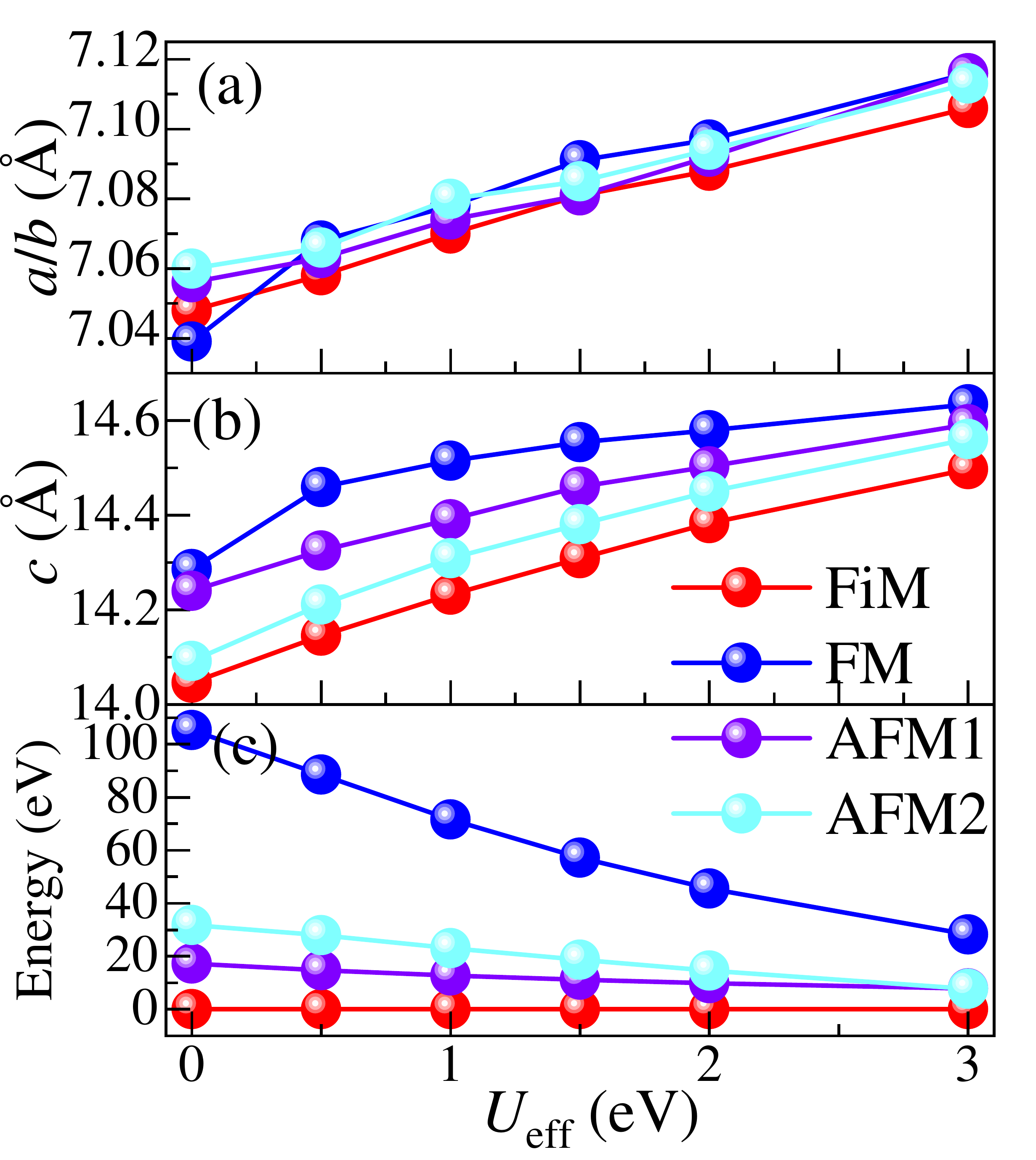}
\caption{DFT results for Mn$_3$Si$_2$Te$_6$ varying $U_{\rm eff}$. (a) Optimized in-plane lattice constants. (b) Optimized $c$-axis lattice constants. (c) Energy (per Mn) of various magnetic orders. The FiM state is taken as the energy of reference.}
\label{U-E}
\end{figure}

Here, all the magnetic orders were fully relaxed based on the LSDA+$U_{\rm eff}$ procedure. First, the in-plane and $c$-axis lattice constants of various magnetic orders for relaxed structures are summarized in Figs.~\ref{U-E}(a) and (b), varying $U_{\rm eff}$. As $U_{\rm eff}$ increases, the calculated lattice constants increases. Clearly, the $U_{\rm eff} = 0.5 $ eV value gives the most accurate structure, where our optimized lattice constants are $a = b = 7.058$, $c = 14.145$~\AA ~for the FiM spin state, close to the low-temperature experimental results ($a = b = 7.017$, $c = 14.172$~\AA~\cite{Ni:prb21}. As shown in Fig.~\ref{U-E}(c), the FiM state always has the lowest energy among all candidate configurations, in agreement with neutron experiments. Increasing $U_{\rm eff}$, the calculated magnetic moment increases from $4.174/4.009$ to $4.478/4.435$ $\mu_{\rm B}$/Mn, for the Mn1 and Mn2 sites, respectively.

\begin{figure}
\centering
\includegraphics[width=0.48\textwidth]{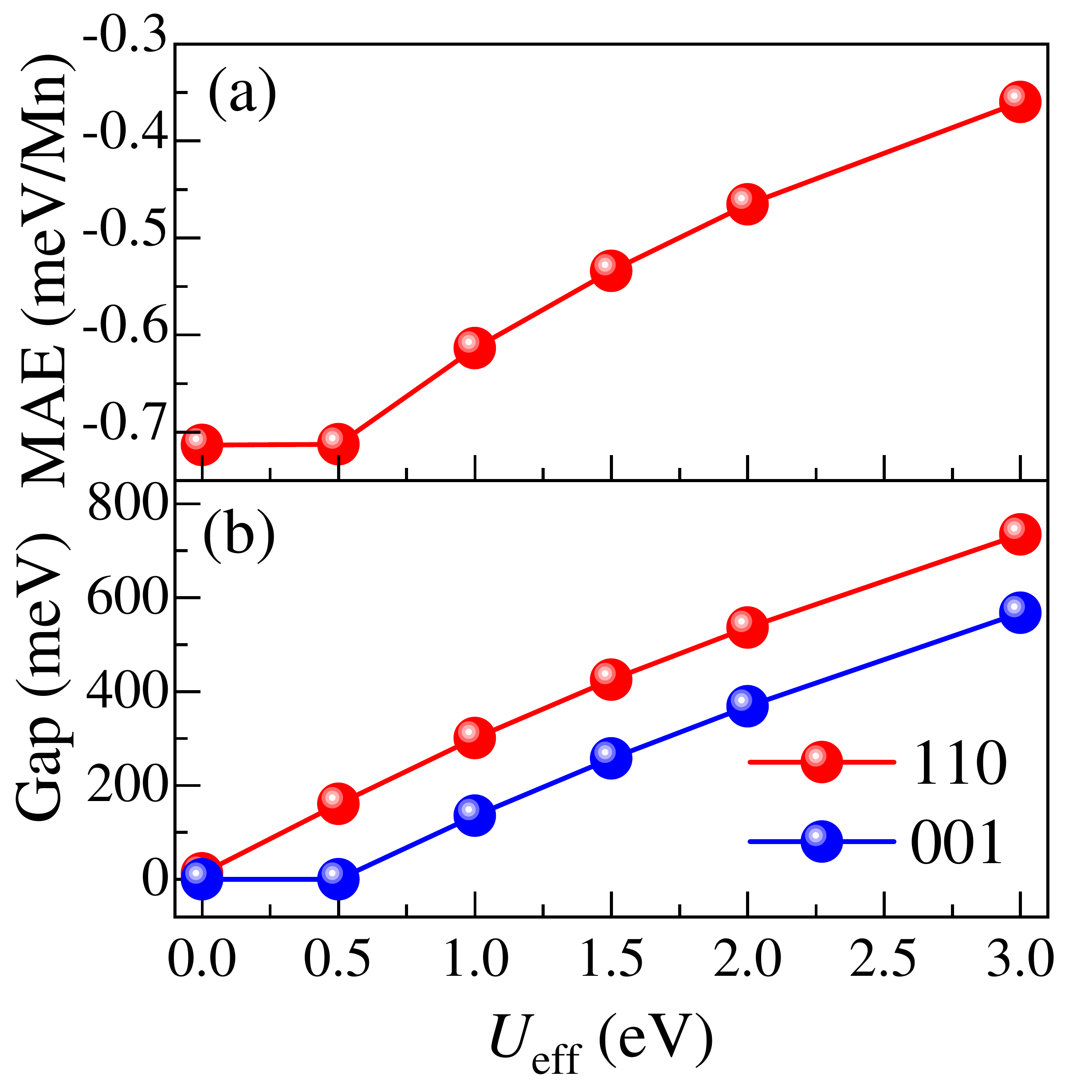}
\caption{The calculated MAE and band gap of Mn$_3$Si$_2$Te$_6$ for the FiM state varying $U_{\rm eff}$. (a) MAE (energy per Mn). (b) Gap (in meV).}
\label{U-MAE}
\end{figure}

Turning on the SOC, the spin quantization axis of the FiM state still points along the $ab$ plane, independently of $U_{\rm eff}$. In addition, we also calculated the MAE (E[110]-E[001]) for the FiM state varying $U_{\rm eff}$, as displayed in Fig.~\ref{U-MAE}(a). As expected, the band gap of the FiM state increases as $U_{\rm eff}$ increases. Furthermore, the calculated band gaps are reduced by changing the spin orientation from the [110] to the [001] directions, as shown in Fig.~\ref{U-MAE}(b).  Increasing the values of $U_{\rm eff}$, the total band gap continues to increase in the ground FiM ($m//ab$) state, since increasing $U_{\rm eff}$ would increase the Mott-gap of Mn $3d$ orbitals and shift the occupied Mn $3d$ state to a lower energy region in this system. By switching the spin orientation of FiM to the out-plane direction ($m//c$), the band gap would be reduced. The reduced values of the gap $\Delta$ ([110]-[001]) do not change much varying $U_{\rm eff}$ because this reduced gap is driven by the SOC of the Te atom by lifting the nodal-line degeneracy, independent of the values of $U_{\rm eff}$.

\begin{references}
\bibitem{Imada:rmp} M. Imada, A. Fujimori, and Y. Tokura, \href{https://doi.org/10.1103/RevModPhys.70.1039}{Rev. Mod. Phys. \textbf{70}, 1039 (1998).}
\bibitem{Moreo:science} A. Moreo, S. Yunoki, and E. Dagotto, \href{https://doi.org/10.1126/science.283.5410.2034}{Science \textbf{283}, 2034 (1999).}
\bibitem{Dagotto:science} E. Dagotto, \href{https://doi.org/10.1126/science.1107559}{Science \textbf{309}, 257 (2005).}
\bibitem{Zhou:rmp} Y. Zhou, K. Kanoda, and T.-K. Ng, \href{https://doi.org/10.1103/RevModPhys.89.025003}{Rev. Mod. Phys. \textbf{89}, 025003 (2017).}
\bibitem{Cao:rpp} G. Cao, and P. Schlottmann, \href{https://doi.org/10.1088/1361-6633/aaa979}{Rep. Prog. Phys. \textbf{81}, 042502 (2018).}
\bibitem{Varignon:nc19} J. Varignon, M. Bibes, and A. Zunger, \href{https://doi.org/10.1038/s41467-019-09698-6}{Nat. Commun. \textbf{10}, 1658 (2019).}
\bibitem{Zhang:prb20} Y. Zhang, L.-F. Lin, W. Hu, A. Moreo, S. Dong, and E. Dagotto, \href{https://doi.org/10.1103/PhysRevB.102.195117}{Phys. Rev. B \textbf{102}, 195117 (2020).}
\bibitem{Takayama:jpsj} T. Takayama, J. Chaloupka, A. Smerald, G. Khaliullin, and H. Takagi, \href{https://doi.org/10.7566/JPSJ.90.062001}{J. Phys. Soc. Jpn. \textbf{90}, 062001 (2021).}
\bibitem{Do:prb22} S.-H. Do, K. Kaneko, R. Kajimoto, K. Kamazawa, M. B. Stone, J. Y. Y. Lin, S. Itoh, T. Masuda, G. D. Samolyuk, E. Dagotto, W. R. Meier, B. C. Sales, H. Miao, and A. D. Christianson, \href{https://doi.org/10.1103/PhysRevB.105.L180403}{Phys. Rev. B \textbf{105}, L180403 (2022).}
\bibitem{Dagotto:rp} E. Dagotto, T. Hotta, and A. Moreo, \href{https://doi.org/10.1016/S0370-1573(00)00121-6}{Phys. Rep. \textbf{344}, 1 (2001).}
\bibitem{Tokura:rpp} Y. Tokura, \href{https://doi.org/10.1088/0034-4885/69/3/R06}{Rep. Prog. Phys. \textbf{66}, 797 (2006).}
\bibitem{Wu:prl01} T. Wu, S. B. Ogale, J. E. Garrison, B. Nagaraj, A. Biswas, Z. Chen, R. L. Greene, R. Ramesh, T. Venkatesan, and A. J. Millis, \href{https://doi.org/10.1103/PhysRevLett.86.5998}{Phys. Rev. Lett. \textbf{86}, 5998 (2001).}
\bibitem{Ward:np} T. Z. Ward, J. D. Budai, Z. Gai, J. Z. Tischler, L. Yin and J. Shen, \href{https://doi.org/10.1038/nphys1419}{Nature. Phys. \textbf{5}, 885 (2009).}
\bibitem{Lin:prl18} H. Lin, H. Liu, L. Lin, S. Dong, H. Chen, Y. Bai, T. Miao, Y. Yu, W. Yu, J. Tang, Y. Zhu, Y. Kou, J. Niu, Z. Cheng, J. Xiao, W. Wang, E. Dagotto, L. Yin, and J. Shen, \href{https://doi.org/10.1103/PhysRevLett.120.267202}{Phys. Rev. Lett. \textbf{120}, 267202 (2018).}
\bibitem{miao:pnas2020} T. Miao, L. Deng, W. Yang, J. Ni, C. Zheng, J. Etheridge, S. Wang, H. Liu, H. Lin, Y. Yu, Q. Shi, P. Cai, Y. Zhu, T. Yang, X. Zhang, X. Gao, C. Xi, M. Tian, X. Wu, H. Xiang, E. Dagotto, L. Yin, and J. Shen, \href{https://doi.org/10.1073/pnas.1920502117}{Proc. Natl. Acad. Sci. USA  \textbf{117}, 7090 (2020).}
\bibitem{Sergienko:prb} I. A. Sergienko and E. Dagotto, \href{https://doi.org/10.1103/PhysRevB.73.094434}{Phys. Rev. B \textbf{73}, 094434 (2006).}
\bibitem{Sergienko:prl} I. A. Sergienko, C. \ifmmode \mbox{\c{S}}\else \c{S}\fi{}en, and E. Dagotto, \href{https://doi.org/10.1103/PhysRevLett.97.227204}{Phys. Rev. Lett. \textbf{97}, 227204 (2006).}
\bibitem{Brink:jpcm} J. v. d. Brink and D. I. Khomskii, \href{https://doi.org/10.1088/0953-8984/20/43/434217}{J. Phys.: Condens. Matter \textbf{20}, 434217 (2008).}
\bibitem{Dong:nsr} S. Dong, H. Xiang, and E. Dagotto, \href{https://doi.org/10.1093/nsr/nwz023}{ Nat. Sci. Rev. \textbf{6}, 629 (2019).}
\bibitem{Balachandran:prb13} P. V. Balachandran and J. M. Rondinelli, \href{https://doi.org/10.1103/PhysRevB.88.054101}{Phys. Rev. B \textbf{88}, 054101 (2013).}
\bibitem{Varignon:prb17} J. Varignon, M. N. Grisolia, D. Preziosi, P. Ghosez, and M. Bibes, \href{https://doi.org/10.1103/PhysRevB.96.235106}{Phys. Rev. B \textbf{96}, 235106 (2017).}
\bibitem{Pandey:prb21} B. Pandey, Y. Zhang, N. Kaushal, R. Soni, L.-F. Lin, W.-J. Hu, G. Alvarez, and E. Dagotto, \href{https://doi.org/10.1103/PhysRevB.103.045115}{Phys. Rev. B \textbf{103}, 045115 (2021).}
\bibitem{Lin:prm21} L.-F. Lin, N. Kaushal, Y. Zhang, A. Moreo, and E. Dagotto, \href{https://doi.org/10.1103/PhysRevMaterials.5.025001}{Phys. Rev. Matter. \textbf{5}, 235106 (2021).}
\bibitem{Dagotto:rmp94} E. Dagotto, \href{https://doi.org/10.1103/RevModPhys.66.763}{Rev. Mod. Phys. \textbf{66}, 763 (1994).}
\bibitem{Dagotto:Rmp} E. Dagotto, \href{https://doi.org/10.1103/RevModPhys.85.849}{Rev. Mod. Phys. \textbf{85}, 849 (2013).}
\bibitem{Dai:rmp} P. Dai, \href{https://doi.org/10.1103/RevModPhys.87.855}{Rev. Mod. Phys. \textbf{87}, 855 (2015).}
\bibitem{Li:nature}  D. Li, K. Lee, B. Y. Wang, M. Osada, S. Crossley, H. R. Lee, Y. Cui, Y. Hikita, and H. Y. Hwang, \href{https://doi.org/10.1038/s41586-019-1496-5}{Nature \textbf{527}, 624 (2019).}
\bibitem{May:prb17} A. F. May, Y. Liu, S. Calder, D. S. Parker, T. Pandey, E. Cakmak, H. Cao, J. Yan, and M. A. McGuire, \href{https://doi.org/10.1103/PhysRevB.95.174440}{Phys. Rev. B \textbf{95}, 174440 (2017).}
\bibitem{Sala:prb22} G. Sala, J. Y. Y. Lin, A. M. Samarakoon, D. S. Parker, A. F. May, and M. B. Stone, \href{https://doi.org/10.1103/PhysRevB.105.214405}{Phys. Rev. B \textbf{105}, 214405 (2022).}
\bibitem{Liu:prb21} Y. Liu, Z. Hu, M. Abeykoon, E. Stavitski, K. Attenkofer, E. D. Bauer, and C. Petrovic, \href{https://doi.org/10.1103/PhysRevB.103.245122}{Phys. Rev. B \textbf{103}, 245122 (2021).}
\bibitem{Ni:prb21} Y. Ni, H. Zhao, Y. Zhang, B. Hu, I. Kimchi, and G. Cao, \href{https://doi.org/10.1103/PhysRevB.103.L161105}{Phys. Rev. B \textbf{103}, L161105 (2021).}
\bibitem{Wang:prb22} J. Wang, S. Wang, X. He, Y. Zhou, C. An, M. Zhang, Y. Zhou, Y. Han, X. Chen, J. Zhou, and Z. Yang, \href{https://doi.org/10.1103/PhysRevB.106.045106}{Phys. Rev. B \textbf{106}, 045106 (2022).}
\bibitem{Zhang:nature} Y. Zhang, Y. Ni, H. Zhao, S. Hakani, F. Ye, L. DeLong, I. Kimchi and G. Cao, \href{https://doi.org/10.1038/s41586-022-05262-3}{Nature (2022).}
\bibitem{Seo:nature} J. Seo, C. De, H. Ha, et al., \href{https://doi.org/10.1038/s41586-021-04028-7}{Nature \textbf{599}, 576 (2021).}
\bibitem{Ye:arxiv} F. Ye, M. Matsuda, Z. Morgan, T. Sherline, Y. Ni, H. Zhao, and G. Cao, \href{https://doi.org/10.1103/PhysRevB.106.L180402}{Phys. Rev. B \textbf{106}, L180402 (2022).}
\bibitem{Mijin:arxiv} S. D. Mijin, A. {\v{S}}olaji{\'c}, J. Pe{\v{s}}i{\'c}, Y. Liu, C. Petrovic, M. Bockstedte, A. Bonanni, Z. V. Popovi{\'c}, and N. Lazarevi{\'c}, Nenad, \href{https://doi.org/10.48550/arXiv.2209.02664}{arXiv:2209.02664v1 (2022).}
\bibitem{Kresse:Prb} G. Kresse and J. Hafner, \href{https://doi.org/10.1103/PhysRevB.47.558}{Phys. Rev. B \textbf{47}, 558 (1993).}
\bibitem{Kresse:Prb96} G.~Kresse and J.~Furthm\"{u}ller, \href{https://doi.org/10.1103/PhysRevB.54.11169}{Phys. Rev. B \textbf{54}, 11169 (1996).}
\bibitem{Blochl:Prb} P. E. Bl\"{o}chl, \href{https://doi.org/10.1103/PhysRevB.50.17953}{Phys. Rev. B \textbf{50}, 17953 (1994).}
\bibitem{Perdew:Prl} J. P. Perdew, K. Burke, and M. Ernzerhof, \href{https://doi.org/10.1103/PhysRevLett.77.3865}{Phys. Rev. Lett. \textbf{77}, 3865 (1996).}
\bibitem{Dudarev:prb} S. L. Dudarev, G. A. Botton, S. Y. Savrasov, C. J. Humphreys, and A. P. Sutton, \href{https://doi.org/10.1103/PhysRevB.57.1505}{Phys. Rev. B \textbf{57}, 1505 (1998).}
\bibitem{Momma:vesta} K. Momma and F. Izumi, \href{https://doi.org/10.1107/S0021889811038970}{J. Appl. Crystallogr. \textbf{44}, 1272 (2011).}
\bibitem{Supplemental} See Supplemental Material at \href{http://link.aps.org/supplemental/10.1103/PhysRevB.xx/xxxxxx}{http://link.aps.org/supplemental/10.1103/PhysRevB.xx/xxxxxx.} for more results.
\bibitem{Perdew:Prl08} J. P. Perdew, A. Ruzsinszky, G. I. Csonka, O. A. Vydrov, G. E. Scuseria, L. A. Constantin, X. Zhou and K. Burke, \href{https://doi.org/10.1103/PhysRevLett.100.136406}{Phys. Rev. Lett. \textbf{100}, 136406 (2008).}
\bibitem{Lin:fop}L.-F. Lin, L. Z. Wu, and S. Dong, \href{https://doi.org/10.1007/s11467-016-0584-3}{Front. Phys. \textbf{11}, 117502 (2016).}
\bibitem{Zhang:prm} Y. Zhang, L.-F. Lin, J.-J. Zhang, X. Huang, M. An, and S. Dong, \href{https://doi.org/10.1103/PhysRevMaterials.1.034406}{Phys. Rev. Mater. \textbf{1}, 034406 (2017).}
\bibitem{Lin:prm17} L.-F. Lin, Q.-R. Xu, Y. Zhang, J.-J. Zhang, Y.-P. Liang, and S. Dong, \href{https://doi.org/10.1103/PhysRevMaterials.1.071401}{Phys. Rev. Materials {\bf 1}, 071401(R) (2017).}
\bibitem{Zhang:prb20-2} Y. Zhang, L.-F. Lin, A. Moreo, S. Dong, and E. Dagotto, \href{https://journals.aps.org/prb/abstract/10.1103/PhysRevB.101.144417}{Phys. Rev. B \textbf{101}, 144417 (2020).}
\bibitem{Salamon:rmp} M. B. Salamon and M. Jaime, \href{https://doi.org/10.1103/RevModPhys.73.583}{Rev. Mod. Phys. \textbf{73}, 583 (2001).}
\bibitem{Liechtensteincontext} In addition, we also studied the effect of the exchange interaction $J$ using the LSDA+$U$+$J$ with the Liechtenstein formulation~\cite{Liechtenstein:prb}. As expected, the magnetic coupling parameters do not change as discussed in the Supplemental Materials.
\bibitem{Liechtenstein:prb} A. I. Liechtenstein, V. I. Anisimov, and J. Zaanen, \href{https://doi.org/10.1103/PhysRevB.52.R5467}{Phys. Rev. B \textbf{52}, R5467 (1995).}
\bibitem{Du:qm} F. Du, L. Yang, Z. Nie, N. Wu, Y. Li, S. Luo, Y. Chen, D. Su, M. Smidman, Y. Shi, C. Cao, F. Steglich, Y. Song and H. Yuan, \href{https://doi.org/10.1038/s41535-022-00468-0}{npj Quantum Mater. \textbf{7}, 65 (2022).}
\bibitem{Bellaiche:Prb} L. Bellaiche and D. Vanderbilt, \href{https://doi.org/10.1103/PhysRevB.61.7877}{Phys. Rev. B \textbf{61}, 7877 (2000).}
\bibitem{Ramer:Prb} N. J. Ramer and A. M. Rappe, \href{https://doi.org/10.1103/PhysRevB.62.R743}{Phys. Rev. B \textbf{62}, R743(R) (2000).}
\bibitem{Zhang:prb21} Y. Zhang, L.-F. Lin, A. Moreo, G. Alvarez, and E. Dagotto \href{https://doi.org/10.1103/PhysRevB.103.L121114}{Phys. Rev. B \textbf{103}, L121114 (2021).}

\end{references}
\end{document}